\documentclass[%
 reprint,
 superscriptaddress,
nofootinbib,
 amsmath,amssymb,
 aps,
]{revtex4-2}

\usepackage{graphicx}
\usepackage{macros_ads}
\usepackage{dcolumn}
\usepackage{bm}
\usepackage{xcolor}
\usepackage{amsmath}
\usepackage{appendix}
\usepackage{hyperref}
\usepackage{xspace}
\usepackage{orcidlink}
\usepackage{ulem}

\newcommand{\tresid}[0]{\delta\mathbf{t}}
\newcommand{\tresidng}[0]{\delta\mathbf{t}^{15\textrm{yr}}}
\newcommand{\fmat}[0]{\mathbf{F}}
\newcommand{\tmat}[0]{\mathbf{T}}
\newcommand{\avec}[0]{\mathbf{a}}
\newcommand{\rvec}[0]{\mathbf{r}}
\newcommand{\mmat}[0]{\mathbf{M}}
\newcommand{\nmat}[0]{\mathbf{N}}
\newcommand{\bvec}[0]{\mathbf{b}}
\newcommand{\bmat}[0]{\mathbf{B}}

\newcommand{\cmat}[0]{\mathbf{C}}

\newcommand{\epsvec}[0]{\boldsymbol{\epsilon}}
\newcommand{\phimat}[0]{\boldsymbol{\varphi}}

\newcommand{\etavec}[0]{\boldsymbol{\eta}}
\newcommand{\rhovec}[0]{\boldsymbol{\rho}}

\newcommand{\curn}[0]{\textsc{curn}\xspace}
\newcommand{\hd}[0]{\textsc{hd}\xspace}
\newcommand{\dd}[0]{\textrm{d}}
\newcommand{\hdposterior}[0]{\textsc{HDPosteriorDraws}\xspace}

\begin{document}
\newcommand{\resultNumber}[1]{#1\xspace}
\newcommand{\pbfFullArray}[0]{\resultNumber{873}}
\newcommand{\pbfFullArrayEnterprise}[0]{\resultNumber{506}}
\newcommand{\pbfSimulationSigmaVal}[0]{\resultNumber{3.0}}
\newcommand{\pbfSimulationSigmaValEnterprise}[0]{\resultNumber{2.9}}

\newcommand{\numSkyScrambles}[0]{\resultNumber{400,000}}
\newcommand{\SkyScrambleBayesianPvalSigma}[0]{\resultNumber{3.7}}
\newcommand{\SkyScrambleBayesianPval}[0]{\resultNumber{1\times10^{-4}}}
\newcommand{\SkyScrambleNormalPval}[0]{\resultNumber{4.2}}

\newcommand{\HDReconstructionMinPval}[0]{\resultNumber{0.40}}
\newcommand{\HDReconstructionMaxPval}[0]{\resultNumber{0.88}}

\title{The NANOGrav 15 yr data set: Posterior predictive checks for gravitational-wave detection with pulsar timing arrays}

\author{ Gabriella Agazie\,\orcidlink{0000-0001-5134-3925}}
\affiliation{Center for Gravitation, Cosmology and Astrophysics, Department of Physics, University of Wisconsin-Milwaukee,\\ P.O. Box 413, Milwaukee, WI 53201, USA}
\author{ Akash Anumarlapudi\,\orcidlink{0000-0002-8935-9882}}
\affiliation{Center for Gravitation, Cosmology and Astrophysics, Department of Physics, University of Wisconsin-Milwaukee,\\ P.O. Box 413, Milwaukee, WI 53201, USA}
\author{ Anne M. Archibald\,\orcidlink{0000-0003-0638-3340}}
\affiliation{Newcastle University, NE1 7RU, UK}
\author{ Zaven Arzoumanian\,\orcidlink{}}
\affiliation{X-Ray Astrophysics Laboratory, NASA Goddard Space Flight Center, Code 662, Greenbelt, MD 20771, USA}
\author{ Jeremy George Baier\,\orcidlink{0000-0002-4972-1525}}
\affiliation{Department of Physics, Oregon State University, Corvallis, OR 97331, USA}
\author{ Paul T. Baker\,\orcidlink{0000-0003-2745-753X}}
\affiliation{Department of Physics and Astronomy, Widener University, One University Place, Chester, PA 19013, USA}
\author{ Laura Blecha\,\orcidlink{0000-0002-2183-1087}}
\affiliation{Physics Department, University of Florida, Gainesville, FL 32611, USA}
\author{ Adam Brazier\,\orcidlink{0000-0001-6341-7178}}
\affiliation{Cornell Center for Astrophysics and Planetary Science and Department of Astronomy, Cornell University, Ithaca, NY 14853, USA}
\affiliation{Cornell Center for Advanced Computing, Cornell University, Ithaca, NY 14853, USA}
\author{ Paul R. Brook\,\orcidlink{0000-0003-3053-6538}}
\affiliation{Institute for Gravitational Wave Astronomy and School of Physics and Astronomy, University of Birmingham, Edgbaston, Birmingham B15 2TT, UK}
\author{ Sarah Burke-Spolaor\,\orcidlink{0000-0003-4052-7838}}
\altaffiliation{Sloan Fellow}
\affiliation{Department of Physics and Astronomy, West Virginia University, P.O. Box 6315, Morgantown, WV 26506, USA}
\affiliation{Center for Gravitational Waves and Cosmology, West Virginia University, Chestnut Ridge Research Building, Morgantown, WV 26505, USA}
% ...

\date{\today}

\begin{abstract}
Pulsar-timing-array experiments have reported evidence for a stochastic background of nanohertz gravitational waves consistent with the signal expected from a population of supermassive--black-hole binaries.
Their analyses assume power-law spectra for intrinsic pulsar noise and for the background, as well as a Hellings--Downs cross-correlation pattern among the gravitational-wave--induced residuals across pulsars.
These assumptions may not be realized in actuality.
We test them in the NANOGrav 15 yr data set using Bayesian posterior predictive checks. After fitting our fiducial model to real data, we generate a population of simulated data-set replications. We use the replications to assess whether the optimal-statistic significance, inter-pulsar correlations, and spectral coefficients are extreme.
We recover Hellings--Downs correlations in simulated data sets at significance levels consistent with the correlations measured in the NANOGrav 15 yr data set.
A similar test on spectral coefficients shows that their values in real data are not extreme compared to their distributions across replications.
We also evaluate the evidence for the stochastic background using posterior-predictive versions of the frequentist optimal statistic and of Bayesian model comparison, and find comparable significance (3.2\ $\sigma$ and 3\ $\sigma$ respectively) to what was previously reported for the standard statistics.
We conclude with novel visualizations of the reconstructed gravitational waveforms that enter the residuals for each pulsar.
Our analysis strengthens confidence in the identification and characterization of the gravitational-wave background.
\end{abstract}

\maketitle

\section{\label{sec:intro}Introduction}

In June 2023, four separate publications based on the observations of five pulsar-timing-array (PTA) collaborations reported strong evidence for a nanohertz gravitational-wave (GW) background~\cite{NANOGrav:2023gor,Reardon:2023gzh,Xu:2023wog,EPTA:2023fyk}, spurring interest in the implications of its spectral properties and spatial correlations for astrophysics and fundamental physics~\cite{EPTA:2023xxk, NANOGrav:2023hvm,NANOGrav:2023hfp,Agazie:2024jbf}.
If the signal originates from a population of supermassive black hole binaries (SMBHBs), its spectrum is expected to approximate a power law~\cite{Phinney:2001di, Sesana:2012ak}, but deviations can be caused by a large number of potential effects. For example,
at low frequencies, interactions between the binaries and the surrounding gas may result in a spectral turnover; at high frequencies, the finite number of binaries emitting in each frequency bin may result in bin-to-bin fluctuations~\cite{NANOGrav:2023hfp,EPTA:2023xxk,NANOGrav:2023ygs}.
If the signal originates from new physics, the spectrum can point to the mechanism of its generation, and a large number of models are currently consistent with the data~\cite{NANOGrav:2023hvm,EuropeanPulsarTimingArray:2023egv}. 

Spatial correlations between pulse times of arrival (TOAs) for different pulsars were found to be consistent with the Hellings--Downs function, the correlation pattern induced by an isotropic GW background~\cite{Sazhin1978,Detweiler1979,hellings_upper_1983}. 
Deviations could be caused by anisotropy in the background, by a signal from a loud individual SMBHB.
Measuring anisotropy would constrain black-hole population properties~\cite{Gardiner:2023zzr}, while detecting an individual SMBHB would offer a prime target for multi-messenger follow up.
However, dedicated searches for anisotropy and individual sources have so far produced null results~\cite{NANOGrav:2023pdq,NANOGrav:2023tcn,EPTA:2023gyr}. Systematic errors could also induce correlations between pulsars, e.g. monopolar correlations due to clock errors, or dipolar correlations induced by errors in the solar system ephemeris~\cite{TiburziHobbs2016,NANOGrav:2020tig_bayesephem}. There is slight evidence for monopolar correlations presented in the NANOGrav 15 yr data set~\cite{NANOGrav:2023gor}.

Simulations can address the expected level of anisotropy from a population of SMBHBs~\cite{Becsy:2022pnr,Gardiner:2023zzr} and its detectability using standard PTA models, which assume an isotropic GW background with Gaussian statistics and a stationary power-law spectrum.
Indeed, Refs.~\cite{Becsy:2023qul,Valtolina:2023axp} found that the GW signal from a realistic SMBHB population would still be detected using standard models.
Thus, current PTA observations~\cite{NANOGrav:2023gor, Reardon:2023gzh, Xu:2023wog,EPTA:2023fyk,InternationalPulsarTimingArray:2023mzf} do not preclude the presence of astrophysically interesting deviations from power-law spectrum or isotropy. 

In this paper, we ask whether the power-law and Hellings--Downs assumptions are supported by observed data, independently of any specific alternative physical model.
Our starting point is a fiducial Bayesian analysis of NANOGrav's 15 yr data set~\cite{NANOGrav:2023hde} under the standard power-law, Hellings--Downs model.
We test these assumptions by way of \textit{posterior predictive model checks}~\cite{GelmanCarlin2014} as proposed in the context of PTA data in Refs.~\cite{ppc1,ppc2}. These checks consist of creating populations of \textit{replicated} data sets from real-data parameter posteriors, and using these replications to evaluate whether real data is ``typical'' (i.e., not a statistical outlier) according to a variety of detection, spectral, and correlation statistics. Similar types of checks are becoming increasingly common in the realm of binary black hole population analyses as well~\cite{LIGOScientific:2018jsj,Fishbach:2019ckx,Fishbach:2020qag,LIGOScientific:2020kqk,KAGRA:2021duu,Essick:2021vlx,Callister:2022qwb,Payne:2022xan,Miller:2024sui}.

Specifically, following Ref.~\cite{ppc1} we re-evaluate the significance of Hellings--Downs correlations and search for alternative spatial correlations using a new detection statistic that marginalizes $p$-values over noise-parameter posteriors.
Following Ref.~\cite{ppc2} we test the power-law assumption by comparing intrinsic-noise and GW power-spectrum posteriors as computed for real and replicated data,
and we perform a similar test for the binned angular correlations between pulsars. 
We also carry out leave-one-out cross validation to identify possible mismodeling in individual pulsars, and to compute the pseudo Bayes factor (a cross-validation metric of model comparison) between the standard Hellings--Downs model and a null model in which common excess power has no inter-pulsar correlations.

The rest of this paper is organized as follows. 
In Sec.~\ref{sec:data_model_replications} we describe our data and data model, and we introduce two sets of data replications that we will use for model checking.
In Sec.~\ref{sec:bayes_pval_os} we test Hellings--Downs correlations using \textit{Bayesian $p$-values} \cite{ppc1} for the optimal statistic \cite{Anholm:2008wy,chamberlinOS}; these $p$-values are marginalized over GW and intrinsic-noise posteriors, and therefore account fairly for the risk of false positives when the null distribution is uncertain.
We find evidence for Hellings--Downs correlations at the $3.2\,\sigma$ level.
We also evaluate the evidence for additional background components with monopolar or dipolar correlations, and find none.

In Sec.~\ref{sec:spectrum_tests} we compare real-data and replicated-data posteriors to search for deviations from a power-law spectrum and from Hellings--Downs correlations. We find no evidence that any individual frequency bin deviates from the power-law model for either intrinsic pulsar noise or the GW background, consistent with Ref.\ \cite{NG15detchar}.
We also find no evidence that any of the binned inter-pulsar correlations deviates from the Hellings--Downs curve.

In Sec.~\ref{sec:leave_one_out} we examine the predictive power of the standard PTA model as fit to the NANOGrav 15 yr data.
We perform a leave-one-out analysis where we fit Hellings--Downs and uncorrelated models to $N_p-1$ pulsars, and use the models to predict the $N_p^{\textrm{th}}$ pulsar's data.
The resulting pseudo Bayes factors favor Hellings--Downs correlations at the $3\,\sigma$ level.
Using simulations, we show that the distribution of the factors across pulsars is consistent with what would be expected for a power-law, Hellings--Downs-correlated GW background with parameters from our fiducial analysis.

Last, in Sec.~\ref{sec:waveforms} we present the gravitational waveforms that can be reconstructed for each pulsar from our fiducial posteriors.
These reconstructions are akin to the waveform reconstructions for stellar-binary coalescences based on LIGO data~\cite{LIGOScientific:2016aoc,LIGOScientific:2018mvr}, with the distinction being that in this case we show the estimated realization of a broadband, spatially correlated stochastic signal, as opposed to the gravitational waveform produced by a single binary system.
Pulsar J1909$-$3744 offers the best view so far of the GW background reported in~\cite{NANOGrav:2023gor, Reardon:2023gzh, Xu:2023wog,EPTA:2023fyk,InternationalPulsarTimingArray:2023mzf}.
In Sec.~\ref{sec:conclusions} we offer concluding remarks.

\section{Data, model, and data replications}
\label{sec:data_model_replications}

In this section we introduce the NANOGrav 15 yr data set, the modeling that is performed on each pulsar, and the full PTA models used to search for a GW background. 
We then discuss data replications based on our typical PTA models, which we use in subsequent sections to compare to the 15 yr data set for the purposes of model checking and model comparison.

\subsection{Data}
\label{ssec:data}

We use the NANOGrav 15 yr data set, which contains 67 pulsars that have been timed for more than 3 years, with 16.03\,yrs of data between the first and the last time of arrival in the data set~\cite{NANOGrav:2023hde}. 
We use the DMX dispersion measure noise model~\cite{Jones:2016fkk} and white noise parameters included in the NANOGrav 15 yr data release~\cite{NANOGrav:2023hde}. 
For each pulsar, a best fit timing model is constructed that accounts for deterministic effects like Roemer delay, proper motion, parallax, binary orbits, etc., which is then subtracted from the TOAs to produce a set of timing residuals for each pulsar, $\tresid$.
Stochastic processes like achromatic intrinsic spin-wandering and GW background-induced delays are included in this initial fit as a single ``total red noise'' contribution, as the first pass analysis is done on a pulsar-by-pulsar basis and so we cannot separate intrinsic pulsar noise from the GW background.

\subsection{PTA model}
\label{ssec:data_model}

In this subsection, we discuss the full PTA model that is used to search for a GW background. 
Readers familiar with this already can skip to Sec.~\ref{sec:bayes_pval_os}, although later we will make frequent reference to equations introduced in this section. For a more in depth presentation of the PTA analyses, see Refs.~\cite{Taylor:2021yjx,vanHaasteren:2012hj,NG9yrGWB}.

The starting point for the analysis are the timing residuals, $\tresid$. 
We characterize stochastic processes like intrinsic pulsar noise and the GW background in the frequency domain using a Fourier matrix $\fmat$ and associated amplitudes $\avec$~\cite{2013PhRvD..87j4021L}. 
The stochastic processes are covariant with elements of the timing model (specifically the frequency, spin-down, and dispersion measure variations), and so we also introduce deviations from the best-fitting timing model parameters, $\epsvec$. 
We assume these deviations are small, such that changes in $\tresid$ are linear in changes in $\epsvec$ with a design matrix $\mmat$ made up of derivatives of $\tresid$ with respect to the timing model parameters. 
Putting these effects together, we have a model for the residuals
\begin{align}
    \rvec = \tresid - \tmat\bvec\,,
\end{align}
where we have consolidated the frequency domain representation and timing model corrections,
\begin{align}
    \tmat &= \begin{bmatrix}
        \mmat & \fmat
    \end{bmatrix},\\
    \bvec &= \begin{bmatrix}
        \epsvec\\ \avec
    \end{bmatrix}\,.
\end{align}

If radio frequency interference is effectively excised and standard pulse profiles are accurate, the resulting noise is dominated by radiometer noise and ``pulse profile jitter'' which is traditionally assumed to be frequency independent and Gaussian.
This leads to a Gaussian likelihood for the timing residuals
\begin{align}
   \ln p(\tresid | \bvec) = -\frac{1}{2}\left[\rvec^T \nmat^{-1} \rvec + \ln\det(2\pi \nmat)\right]\,,
\end{align}
where the covariance matrix $\nmat$ describes the measurement noise of the individual observations, and is block-diagonal. 
TOAs at different radio frequencies from the same individual observation are correlated with one another due to pulse profile jitter~\cite{NANOGrav:2015qfw}, but TOAs from different observations are uncorrelated.

We assume that the GW background and the intrinsic pulsar noise are stationary, and so they can be characterized by the power spectrum of the GW background, correlations between pulsars, and the power spectrum of the intrinsic pulsar noise in each pulsar. 
The assumption of stationarity for the GW background should hold if the dominant contribution to the background is an ensemble of SMBHBs emitting at roughly constant frequencies. 
The assumption that intrinsic pulsar noise is stationary is one of expedience that should be tested. 
Tests on the European Pulsar Timing Array second data release show no signs of non stationarity~\cite{Falxa:2024rkl}.

Information about the power-law amplitude and spectral index for the intrinsic pulsar noise and the GW background is encoded in the covariance matrix of the sine and cosine amplitudes $\avec$ across pulsars. 
We introduce a set of hyperparameters $\bm\Lambda$ to characterize these power laws. 
We place a Gaussian prior on $\bvec$,
\begin{align}
    \label{eq:model_parameters_hyper_prior}
    \ln p(\bvec | \bm\Lambda) &= -\frac{1}{2}\left[\bvec^T \bmat^{-1} \bvec + \ln\det (2\pi \bmat)\right],\\
    \label{eq:big_b_matrix_definition}
    \textrm{where } \bmat &= \begin{bmatrix}
        \infty & 0\\
        0 & \phimat(\bm\Lambda)
    \end{bmatrix}\,.
\end{align}
We use an improper uniform prior on $\epsvec$ so that its posterior is determined by the likelihood.
This prior is now broadcast across $\bvec$ parameters for each pulsar. 
The covariance matrix of the $\avec$ coefficients is given by $\phimat(\bm\Lambda)$, which contains blocks corresponding to correlations of the Fourier modes between pulsars. 
Diagonal blocks encode information about the power spectrum of the total red noise for a given pulsar, including the intrinsic pulsar noise, $\etavec_{a}(\bm\Lambda)$ (where the $a$ subscript labels the pulsar) and the GW background spectrum $\rhovec(\bm\Lambda)$.
Off-diagonal blocks between pulsars $a$ and $b$ contain (scaled) contributions from the GW background. 
Putting all of this together, the covariance matrix for $\avec$ is
\begin{align}
    \varphi(\bm\Lambda)_{(ai,bj)} = \Gamma_{ab}\rho_i^2(\bm\Lambda)\delta_{ij} + \eta_{ai}^2(\bm\Lambda)\delta_{ij}\delta_{ab}\,,
\end{align}
where $i$ and $j$ label frequencies and $\Gamma_{ab}$ corresponds to the correlations between pulsars. 
Different angular correlation patterns correspond to different models. 
In this paper we consider four models. The first states that  $\Gamma_{ab}$ follows the Hellings--Downs curve (\hd model) that is expected from an isotropic GW background,
\begin{align}
   \label{eq:hd}
    \Gamma_{ab}&= \frac{1}{2}\delta_{ab} + \frac{1}{2} - \frac{\zeta_{ab}}{4} +\frac{3}{2}\zeta_{ab}\ln\zeta_{ab}\,,\\
    \zeta_{ab} &= \frac{1 - \cos \theta_{ab}}{2}\,,
\end{align}
where $\theta_{ab}$ is the angle between pulsars $a$ and $b$ on the sky. 
The second is that $\Gamma_{ab} = \delta_{ab}$, which we call the common uncorrelated red noise, \curn, model.
We will also consider a \textsc{mono} model that is characterized by monopolar correlations, $\Gamma_{ab}=1$ and a model with dipolar correlations, \textsc{dip}, with $\Gamma_{ab}=\cos\theta_{ab}$.
Theoretical models indicate that $\rho_i(\bm\Lambda)$ will roughly take the form of a power law, and past empirical studies suggest that $\eta_{ai}(\bm\Lambda)$ often follows a power-law as well,
\begin{align}
    \eta_{ai}^2(\bm\Lambda) &= \frac{A_{\textrm{rn,a}}^2}{12\pi^2}\left(\frac{f_i}{f_{\textrm{yr}}}\right)^{-\gamma_{\textrm{rn,a}}} \frac{f_{\textrm{yr}}^{-3}}{T}\,,\\
    \rho_i^2(\bm\Lambda) &= \frac{A_{\textrm{gw}}^2}{12\pi^2}\left(\frac{f_i}{f_{\textrm{yr}}}\right)^{-\gamma_{\textrm{gw}}} \frac{f_{\textrm{yr}}^{-3}}{T}\,,
\end{align}
where $A_{\textrm{gw}}$ is the amplitude of the GW background at $f_{\textrm{yr}} = (1 \textrm{yr})^{-1}$, $\gamma_{\textrm{gw}}$ is the negative spectral index, $A_{\textrm{rn,a}}$ is the amplitude of intrinsic pulsar noise for pulsar $a$ and $\gamma_{\textrm{rn,a}}$ its associated spectral index. 
The frequency is given by $f_i=i / T$, and $T$ is the time between the first and last TOAs in the data set. 
For intrinsic pulsar noise we use 30 frequencies, $i\in[1, 30]$ and for the GW background we use $i\in [1, 14]$. 
These numbers were chosen based on individual pulsar fitting (for the intrinsic pulsar noise), and a dedicated \curn analysis that allows for the common spectrum to ``flatten'' at high frequencies, where it then becomes indistinguishable from white noise.

The power-law models for the GW background and intrinsic pulsar noise spectra have amplitudes and spectral indices associated with them:  $A_{\textrm{gw}}$, $\gamma_{\textrm{gw}}$ for the GW background and, $A_{\textrm{rn,a}}$, and $\gamma_{\textrm{rn,a}}$ for each of the $N_p$ pulsars in the array. 
We collectively denote these parameters as $\bm\Lambda$.
To reduce the total number of parameters we need to infer, we typically marginalize over the model parameters $\bvec$, leaving a posterior on the hyperparameters
\begin{align}
    \label{eq:marginalized_likelihood_integral}
    p(\bm\Lambda | \tresid) &= \int \textrm{d}\bvec\, p(\tresid | \bvec) p(\bvec | \bm\Lambda) p(\bm\Lambda)\,,\\
    &=\frac{p(\bm\Lambda)}{\sqrt{\det (2\pi \cmat)}} \exp\left(-\frac{1}{2} \tresid^{T} \cmat^{-1} \tresid\right)\,.\label{eq:full_posterior_on_hyperparameters}
\end{align}
The covariance matrix is now $\cmat = \left(\nmat + \tmat \bmat \tmat^T\right)$, and we introduced a prior on the hyperparameters $p(\bm\Lambda)$. 
We also note that $p(\bvec | \tresid, \bm\Lambda) \propto p(\tresid | \bvec) p(\bvec | \bm\Lambda)=\mathcal N(\widehat \bvec, \bm\Sigma)$, which is normal with mean and covariance given by
\begin{align}
    \label{eq:bvec_max_likelihood}
    \widehat\bvec &= \bm\Sigma \tmat^T\nmat^{-1}\tresid\,, \\
    \label{eq:bvec_covariance_matrix}
    \bm\Sigma &= \left(\tmat^T\nmat^{-1}\tmat + \bmat^{-1}\right)^{-1}\,.
\end{align}

We estimate the marginalized posterior on $\bm\Lambda$ using stochastic sampling methods~\cite{NANOGrav:2023icp} because of the large dimension of $\bm\Lambda$ ($2N_p + 2$ in the case described above). This yields $N_{s}$ samples $\{\bm\Lambda^s\}_{s=1}^{N_s}$ approximately drawn from the posterior,
\begin{align}
    \label{eq:posterior_draws}
    \bm\Lambda^s \sim p(\bm\Lambda | \tresid)\,.
\end{align}

\subsection{Data replications}
\label{ssec:data_replications}
Below we use $\tresid$ to refer to generic timing residuals, $\tresidng$ to refer to residuals from the the 15 yr data set, and $\tresid^{\textrm{rep}}$ to refer to data replications.
We use two models to create sets of data replications to compare to the collected data. 
Each method proceeds along similar lines:
\begin{enumerate}
    \item Choose $\bm\Lambda$ by drawing randomly from $p(\bm\Lambda | \tresidng)$.
    \item Draw $\bvec \sim p(\bvec | \bm\Lambda)$. The choice of $p(\bvec | \bm\Lambda)$ depends upon the set of replications we are performing. We specify details below when we discuss individual replication sets. This method nominally calls for us to draw from the improper prior on $\epsvec$, yielding unusable timing residuals. Therefore, we do not simulate timing model variations, and fix $\epsvec\approx 0$.
    \item Draw $\tresid^{\textrm{rep}} \sim \mathcal{N}(\tmat\bvec, \nmat)$ where $\bvec$ comes from the previous step.
\end{enumerate}

The data replications use different models at each stage. We outline the different data replication sets, their purpose, and what models they use to carry out the procedure described above.
\begin{itemize}
    \item \textsc{CURNPosteriorDraws}: We create simulated data sets based on the \curn model which we index with $s$. 
    We draw $\bm\Lambda^s \sim p(\bm\Lambda |\tresidng,\curn)$, and $\bvec^s \sim \mathcal N(0, \bmat(\bm\Lambda^s) |\curn)$, Eqs.~\eqref{eq:model_parameters_hyper_prior} and~\eqref{eq:big_b_matrix_definition}. 
    The conditioning on \curn implies no correlations between pulsars, $\Gamma_{ab}=\delta_{ab}$. 
    We do not simulate timing model variations, i.e., $\epsvec = 0$. 
    These sets of data replications are compared to the data and recovered model parameters.
    \item \textsc{HDPosteriorDraws}: We create simulated data sets based on the \hd model which we index with $s$. 
    We draw $\bm\Lambda^s \sim p(\bm\Lambda |\tresidng,\hd)$, and $\bvec^s \sim \mathcal N(0, \bmat(\bm\Lambda^s) |\hd)$. 
    The conditioning on \hd implies we include Hellings--Downs correlations between pulsars during simulation.  
    We do not simulate timing model variations, i.e., $\epsvec = 0$.
    These sets of data replications are compared to the real data and recovered model parameters to assess how consistent the data are with the \hd model.
\end{itemize}

\section{\label{sec:bayes_pval_os}Posterior predictive null hypothesis testing}

Given $\bm\Lambda^s \sim p(\bm\Lambda | \tresid)$, we perform \textit{posterior predictive checks} by checking whether specific desired properties of the model are consistent in the data. 
To do this, we construct a test statistic, $T(\tresid, \bm\Lambda)$ that is sensitive to the property we are interested in, and we compare that test statistic calculated in the 15 yr data to the same statistic calculated over data replications. 
Using this method we check (1) whether the 15 yr data are consistent with the lack of correlations assumed by the \curn~model; (2) whether the 15 yr data have correlations that are consistent with the HD curve; and (3) whether the 15 yr data show evidence for alternative spatial correlations, e.g., monopolar or dipolar, inconsistent with both the \hd~model and the \curn~model.

\subsection{\textsc{CURN} model tests}

The \textsc{curn} model is characterized by a lack of spatial correlations between pulsars, $\Gamma_{ab}=\delta_{ab}$. 
To reject this model, we use the optimal statistic signal-to-noise ratio (SNR) as our test statistic~\cite{chamberlinOS,Sardesai:2023qsw,Anholm:2008wy,Vigeland:2018ipb}. The SNR, for a given choice of noise parameters, is distributed according to a generalized $\chi^2$ distribution~\cite{Hazboun:2023tiq}; it is large when Hellings--Downs correlations are present and centered around zero when no spatial correlations are present. 

The optimal statistic depends upon the total red noise in each pulsar (intrinsic pulsar noise and GW background), which we do not know \textit{a priori}.
Therefore, current analyses average the SNR over the posterior distribution on the noise parameters,  
\begin{align}
    \overline{\textrm{SNR}} = \int \textrm{d}\bm\Lambda\,p(\bm\Lambda | \tresidng) \textrm{SNR}(\tresidng;\bm\Lambda)\nonumber\\
    \approx \frac{1}{N_s}\sum_{s=1}^{N_s} \textrm{SNR}(\tresidng; \bm\Lambda^s)\,,
\end{align}
where in the second line we perform a Monte Carlo integral using a finite set of $N_s$ posteriors samples~\cite{Vigeland:2018ipb}. $\overline{\textrm{SNR}}$ is then used as the test statistic for null hypothesis testing. 
The motivation for using this ``noise marginalized optimal statistic'' is that we are marginalizing over uncertainty in the noise and signal parameters when we calculate statistical significance.
One uses ``phase shifts''~\cite{Taylor:2016gpq}, ``sky scrambles''~\cite{Cornish:2015ikx}, or data replications to estimate the null distribution of $\overline{\textrm{SNR}}$ and calculate a $p$-value for the measured $\overline{\textrm{SNR}}$. 
In Ref.~\cite{NANOGrav:2023gor}, $\overline{\textrm{SNR}}\approx 5$, which falls at the $3.5-4\sigma$ level in the null distributions, indicating that the \curn model does not fully describe the data.

Here, we use a more conservative statistic that gives more weight to low-SNR outliers and lower weight to high SNR outliers than the noise marginalized optimal statistic~\cite{ppc1}.
The cumulative distribution function is not linear in the SNR, and so we first calculate the $p$-value of the SNR calculated on each $\bm\Lambda^s$, and then average those $p$-values together.
The resulting $p$-value will be less significant than the one calculated on $\overline{\textrm{SNR}}$.
Conceptually, this can be thought of as averaging over the risk of rejecting the null hypothesis by placing more weight on the most conservative noise realizations. 
By contrast, calculating a $p$-value on $\overline{\textrm{SNR}}$ weighs high-SNR (and therefore less conservative noise realizations) equally to low SNR noise realizations.

We compare the value of the optimal statistic on the observed data to its value on data replications from the posterior predictive distribution. 
We calculate a $p$-value on each $\textrm{SNR}(\tresidng;\bm\Lambda^s)$, and average those $p$-values.
This final, averaged $p$-value is referred to as a \textit{posterior predictive $p$-value} or a \textit{Bayesian $p$-value} because it is marginalized over the posterior predictive distribution for the data~\cite{gelman1996posterior,gelman2013_pval_dists,GelmanCarlin2014}.
We can do this generically, when we do not know the distribution of the test statistic, by calculating
\begin{align}\nonumber
    p_B = \int \int &\Theta\left[\mathrm{SNR}(\tresid^\textrm{rep}, \bm\Lambda)-\mathrm{SNR}(\tresidng,\bm\Lambda)\right]\\
    &\times p(\tresid^{\textrm{rep}}|\bm\Lambda, \textsc{curn})
    \, p(\bm\Lambda | \tresidng) \,\mathrm{d}(\tresid^{\mathrm{rep}} )\,\mathrm{d}\bm\Lambda\,.\label{eq:bayesian_pval}
\end{align}
Here $\Theta$ is the Heaviside function, and $p(\tresid^{\mathrm{rep}}|\bm\Lambda, \textsc{curn})$ could be one of the data replications described in Sec.~\ref{ssec:data_replications}, or it could be a set of ``bootstrapped'' data replications like sky scrambles or phase shifts.\footnote{We use $\tresid^{\mathrm{rep}}$ to also denote sky scrambled and phase shifted data sets, in addition to actual simulated data sets. This is somewhat poor notation. When performing sky scrambles and phase shifts the timing residuals themselves are the same as $\tresidng$ and it is the sky position (sky scrambles) or $\fmat$ (phase shifts) that changes. Nevertheless, we use $\tresid^{\mathrm{rep}}$ to refer generically to any data sets or schemes used to construct the null distribution, for simplicity.}
If the analytic distribution for the SNR for a given choice of $\bm\Lambda^s$ is known, then we do not need to actually perform data replications, and $\int \Theta[\cdot]p(\tresid^{\mathrm{rep}}|\bm\Lambda,\curn)\,\mathrm{d}\tresid^{\mathrm{rep}}$ is the inverse cumulative distribution function for the SNR.
The Bayesian $p$-value reduces to
\begin{align}
    \nonumber p_B(\tresidng) &= \int P[\textrm{SNR}(\tresid^\textrm{rep};\bm\Lambda) > \textrm{SNR}(\tresidng;\bm\Lambda)]\\&\times p(\bm\Lambda | \tresidng)\,\mathrm{d}\bm\Lambda \nonumber\\
    \approx \frac{1}{N_s}&\sum_{s=1}^{N_s} P[\textrm{SNR}(\tresid^\textrm{rep,s}\bm;\bm\Lambda^s) > \textrm{SNR}(\tresidng;\bm\Lambda^s)]\,,
    \label{eq:os_snr_bayes_pval_numerical}
\end{align}
where in the second line we evaluate the integral numerically using draws from $p(\bm\Lambda | \tresidng)$. 
The superscript ``rep'' indicates that the inverse cumulative distribution function on the measured $\textrm{SNR}(\tresidng;\bm\Lambda)$ is calculated over (theoretical or actual) data replications or sky scrambles.

\begin{figure}
    \centering
    \includegraphics[width=\columnwidth]{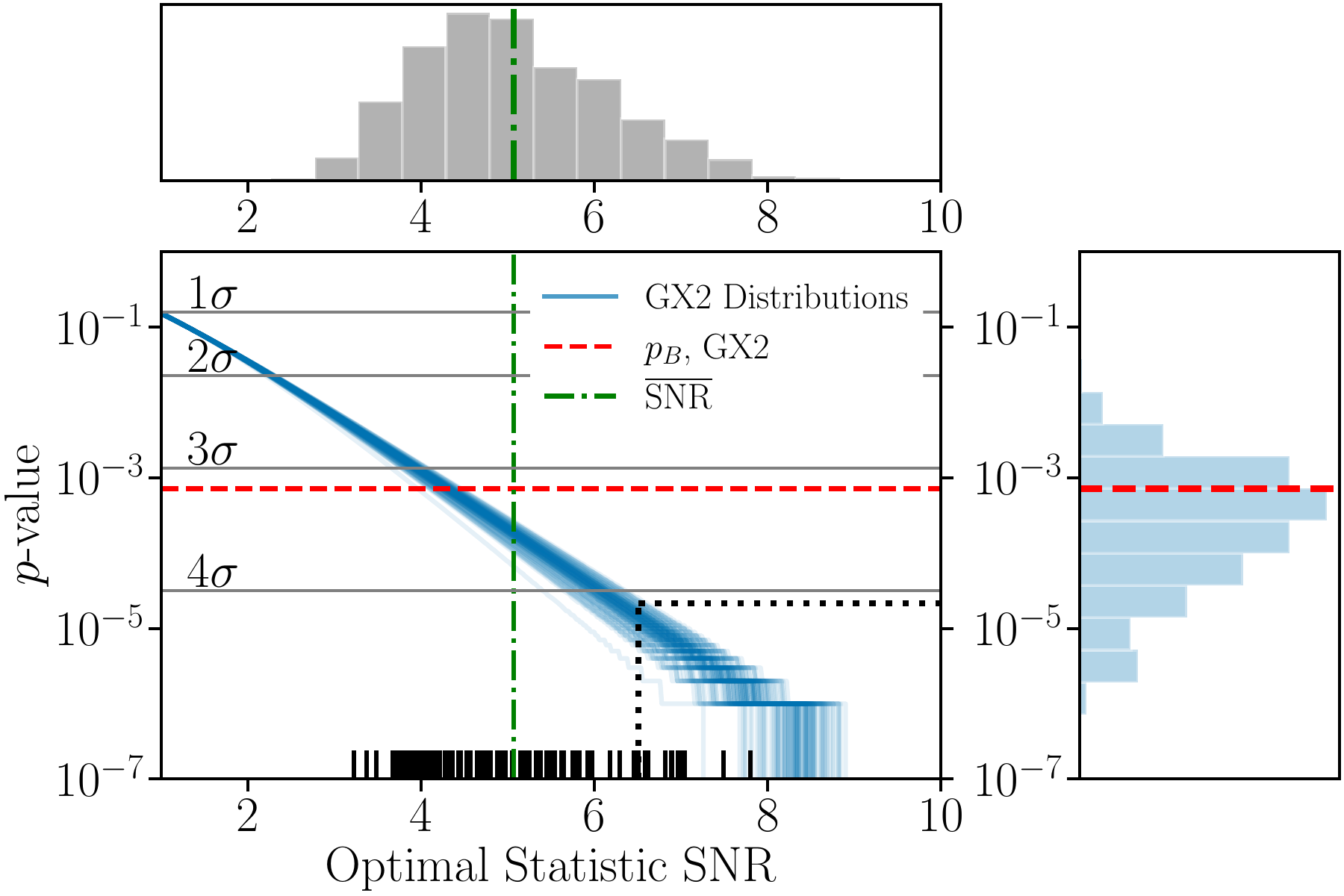}
    \caption{Null hypothesis testing results using the GX2 distribution. The inverse CDF curve is shown in blue for 100 draws from a \curn MCMC chain. The optimal statistic SNR for each of those draws are indicated by the black lines at the bottom, and histogrammed in black in the top panel. We show $p_B$, averaged over $p$-values calculated using the GX2 distribution, with the red dashed line. The histogram of the $p$-value calculated for each draw is shown in the blue histogram in the right panel. The green dot-dashed line indicates $\overline{\textrm{SNR}}$. The gray horizontal lines correspond to different Gaussian equivalent sigma levels.}
    \label{fig:os_bayes_pval_gx2}
\end{figure}

The probability distribution function for the optimal statistic SNR for fixed $\bm\Lambda^s$, under the noise model, is a generalized $\chi^2$ distribution~\cite{Hazboun:2023tiq}
which we will refer to as GX2 moving forward.
In Fig.~\ref{fig:os_bayes_pval_gx2} we show $\textrm{SNR}(\tresidng;\bm\Lambda^s)$ for 100 draws along the bottom, and each blue curve is $P[\textrm{SNR}^{\textrm{rep,s}}(\tresid^\textrm{rep,s}\bm;\bm\Lambda^s) > \textrm{SNR}(\tresidng;\bm\Lambda^s)]$, calculated using the GX2 distribution.
The dashed red line gives $p_B = 7\times10^{-4}$, which corresponds to $3.2\,\sigma$ significance in favor of rejecting the \textsc{curn} model. 
By contrast, the significance of the SNR maximum-likelihood draw from $p(\bm\Lambda|\tresid)$ calculated using GX2 was $\approx 2\times 10^{-4}$ or $3.5\,\sigma$.

\begin{figure}
    \centering
    \includegraphics[width=\columnwidth]{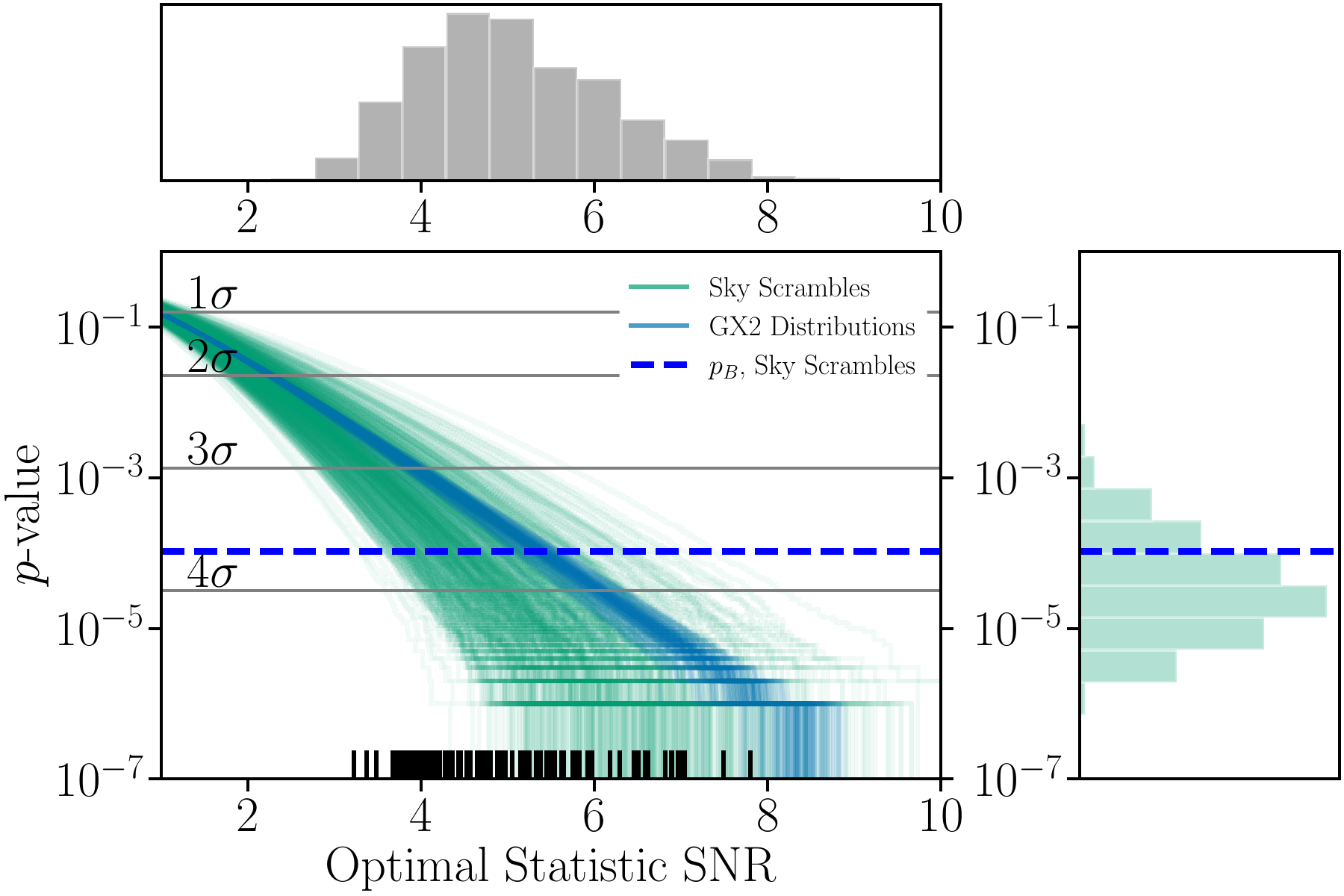}
    \caption{Same as Fig.~\ref{fig:os_bayes_pval_gx2}, but for sky scrambled distributions. Each green line corresponds to an inverse CDF for a given $\bm\Lambda^s$. The green histogram in the right panel corresponds to the $p$-value of the $\mathrm{SNR}(\tresidng;\bm\Lambda^s)$ for each $\bm\Lambda^s$, and the blue dashed line in the middle and right panels corresponds to the average of those $p$-values, which is $p_B$ for sky scrambles. In the center panel, we have included the blue GX2 inverse CDFs from Fig.~\ref{fig:os_bayes_pval_gx2} for reference. The sky scrambles result in a more significant $p$-value and $p_B$.}
    \label{fig:os_bayes_pval_scrambles}
\end{figure}

In using the GX2 distribution, we assumed that the noise model is correct. 
Instead, we can construct replications of our data set that break correlations due to GWs, but preserve any mismodeling in the noise that might cause large SNR values in favor of correlations. 
The two main methods for doing this are sky scrambles~\cite{Cornish:2015ikx} and phase shifts~\cite{Taylor:2016gpq}. 
For each $\bm\Lambda^s$, we perform \numSkyScrambles sky scrambles, where we artificially move the location of the pulsars to different positions on the sky drawn uniformly on the two-sphere and calculate a ``new'' Hellings--Downs curve using these new positions.  
Using these sky scrambles, we build a null distribution and calculate significance.
We repeat this for 100 draws of $\bm\Lambda^s$ and average the inverse CDFs as in Eq.~\eqref{eq:os_snr_bayes_pval_numerical}. 
Under this procedure, we find $p_B=\SkyScrambleBayesianPval$, which corresponds to an equivalent Gaussian significance of $\SkyScrambleBayesianPvalSigma\,\sigma$. 
Using sky scrambles, Ref.~\cite{NANOGrav:2023gor} found $p=5\times 10^{-5}$ using the traditional procedure of building a null distribution for $\overline{\textrm{SNR}}$.
The results are shown in Fig.~\ref{fig:os_bayes_pval_scrambles}, where the inverse CDFs for sky scrambles (green) in general fall off faster than for the GX2. 
However, there are some outliers resulting in $p_B$ being larger than the $p$-value calculated on $\overline{\mathrm{SNR}}$ using scrambles.

It is unclear why the inverse CDF for sky scrambles generally falls off faster than for the GX2; this is an open area of investigation~\cite{DiMarco:2024irz}. In previous work, methods of generating a background distribution from sky scrambling or phase shifting use a ``match statistic,'' in an attempt to use scrambles or shifts that are quasi-independent of one another and the Hellings--Downs curve~\cite{Sampson:2015ada,Taylor:2016ftv}.  Recently, in Ref.~\cite{DiMarco:2023jqq}, the authors suggested only sky scrambles that produce correlation curves that are independent of one another should be used, where independence is achieved by insisting  the match statistic disappear. In Ref.~\cite{NANOGrav:2023gor}, the condition is that the match threshold between any one sky scramble and all others is $\lesssim 0.2$. 
Here we do not use a match statistic, as our goal is to estimate the probability that the pulsars would be arranged on the sky in such a way that noise fluctuations would produce Hellings--Downs correlations. 
To test this, we must draw the positions uniformly on the sky. 
How to produce reliable null distributions for data sets that preserve potentially unmodeled noise is still subject to exploration.

\subsection{Consistency with the \textsc{HD} model}

\begin{figure}
    \centering
    \includegraphics[width=
\columnwidth]{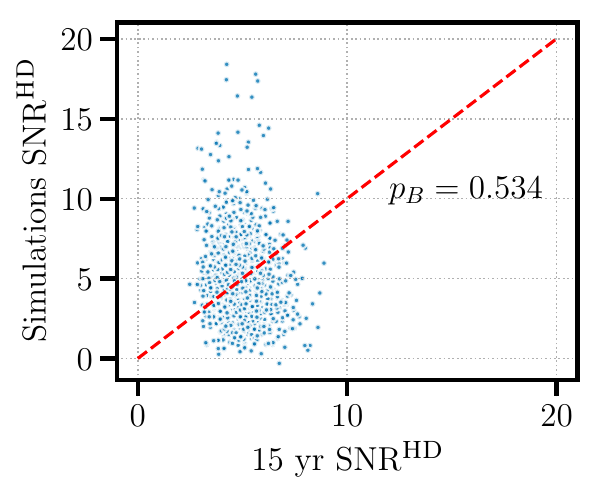}
    \caption{We show a comparison of SNR$(\tresidng, \bm\Lambda^s)$ ($x$-axis) with SNR$(\tresid^{\mathrm{rep}}, \bm\Lambda^s)$ ($y$-axis). Each point corresponds to a draw from the posterior $\bm\Lambda^s$. The fact that roughly half the points fall above the line $y=x$, and $p_B=0.53$ indicates that the measured SNR for Hellings--Downs correlations on the 15 yr data is consistent the 1,000 \hdposterior replications.}
    \label{fig:hd_bayes_pval_hd_only_os}
\end{figure}

In the previous subsection, we reject the null hypothesis of the \curn model at the 3.2\,$\sigma$ level. 
In this section, we use the same scheme to test whether the data are consistent with Hellings--Downs correlations.
Given that we only have an analytic \textit{null} distribution for the optimal statistic, we use a Monte Carlo integral for Eq.~\eqref{eq:bayesian_pval} to evaluate $p_B$ in the presence of a potential signal:
\begin{align}
    p_B \approx \frac{1}{N} \sum_{s=1}^N  \Theta\left[\textrm{SNR}(\tresid^{\textrm{rep}},\bm\Lambda^s)-\textrm{SNR}(\tresidng,\bm\Lambda^s)\right],\label{eq:bayesian_pval_double_sum_approx}
\end{align}
where we average over $N=1,000$ \hdposterior replications.
We show a scatter plot in Fig.~\ref{fig:hd_bayes_pval_hd_only_os}, where the $x$-axis shows $\mathrm{SNR}(\tresid, \bm\Lambda^s)$ and the $y$-axis shows $\mathrm{SNR}(\tresid^{\textrm{rep, s}}, \bm\Lambda^s)$, and $p_B$ corresponds to the fraction of points above the line $y=x$. 
In this case, we find $p_B=0.534$, indicating that the 15 yr NANOGrav data are consistent with data replications that assume Hellings--Downs correlations.

\subsection{Additional spatial correlations}

The model for a GW background assumes spatial correlations that follow the HD curve, but other spatial correlations could arise either from statistical fluctuations or due to mismodeling.
Monopolar correlations could arise due to an error in the clock at each site, corresponding to a correlated offset that is common to all pulsars. 
Dipolar correlations could arise due to an error in the effective location and motion of the solar system barycenter~\cite{TiburziHobbs2016}. 
We use the multiple component optimal statistic~\cite{Sardesai:2023qsw}, which estimates the amplitude of monopolar, dipolar, and Hellings--Downs correlations simultaneously, to test whether our estimate of these correlations in the 15 yr data set is consistent with data replications from a pure HD model. 
The results and methods here follow App. H of Ref.~\cite{NANOGrav:2023gor}. The analysis is nearly identical, but with more simulations used to calculate $p_B$. The results and conclusion are the same as that analysis, but we include it here both for completeness, and because it is strongly related to the rest of the new tests we have performed in this section.

The multiple component optimal statistic \textit{simultaneously} produces the SNR for all three spatial correlations, SNR$^\textrm{MONO}_{\textrm{MC}}$, SNR$^{\textrm{DIP}}_{\textrm{MC}}$, and SNR$^{\textrm{HD}}_{\textrm{MC}}$ where the superscript corresponds to the spatial correlation and the subscript indicates that we are using the multiple component optimal statistic. We again use 1,000 \hdposterior data replications described in Sec.~\ref{sec:data_model_replications} and calculate Eq.~\eqref{eq:bayesian_pval_double_sum_approx} substituting $\textrm{SNR}^\textrm{MONO}_{\textrm{MC}}$ for SNR to produce $p_B^{\textrm{MONO}}$. 
We produce
$p_B^{\textrm{DIP}}$ and $p_B^{\textrm{HD}}$ for dipole and Hellings--Downs correlations defined analogously.

\begin{figure*}
    \includegraphics[width=0.3\textwidth]{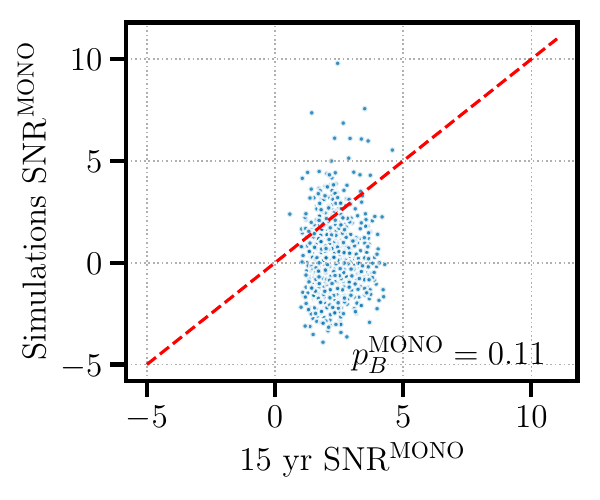}
    \includegraphics[width=0.3\textwidth]{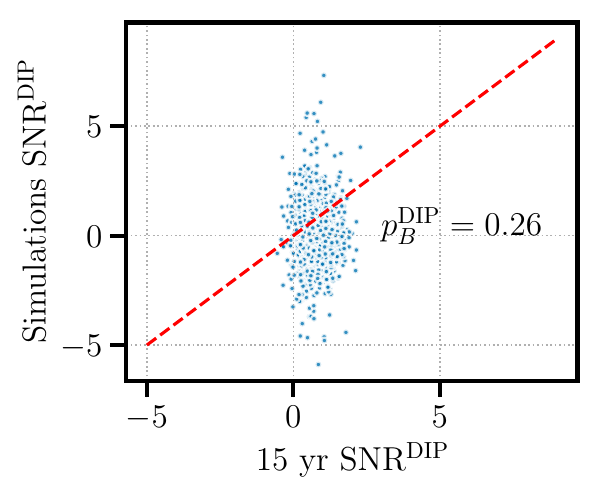}
    \includegraphics[width=0.3\textwidth]{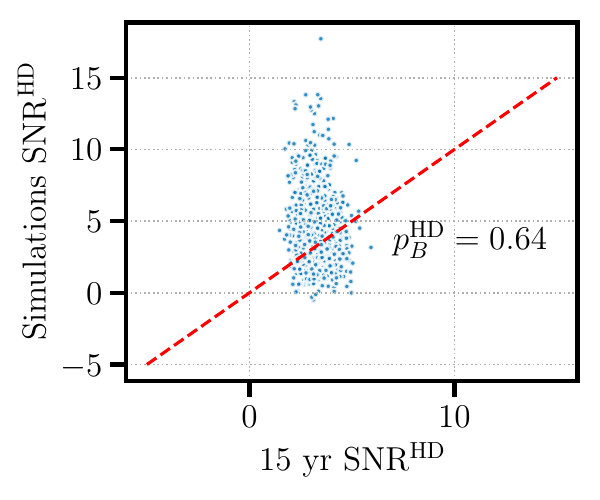}
    \caption{We compare $\textrm{SNR}^{\textrm{\textsc{mono}}}(\tresidng, \bm\Lambda^s)$ ($x$-axis) with $\textrm{SNR}^{\textrm{\textsc{mono}}}(\tresid^{\mathrm{rep}}, \bm\Lambda^s)$ ($y$-axis) in the left panel (\textsc{dip} and \textsc{hd} in center and right panels respectively), using the \hdposterior data replications.
    There is broad consistency between $\textrm{SNR}^{\textsc{hd}}$ in replications and 15 yr data set. For the \textsc{dip} and \textsc{mono} models we find that the recovered SNR is consistent with data replications that include only Hellings--Downs correlations.}
    \label{fig:mcos_bayesian_pvals}
\end{figure*}

We show similar visualizations to the previous section in Fig.~\ref{fig:mcos_bayesian_pvals}. 
We find $p_B^{\textrm{HD}}=0.64$, which again indicates that the HD SNR calculated on the 15 yr data is consistent with what we expect from the pure \hd model. 
Likewise, we find $p_B^{\textrm{DIP}}=0.26$, consistent with no dipolar correlations. 
Finally, we find $p_B^{\textrm{MONO}}=0.11$, which is largely consistent with no monopolar correlations.

\section{Testing spectrum and correlation models}
\label{sec:spectrum_tests}

In this section, we assess the power-law assumption for the GW background and the intrinsic pulsar noise. 
We recap how to estimate the posterior distribution on the Fourier coefficients for the red noise, $\avec$ at each frequency and for each pulsar for both the intrinsic pulsar noise and the GW background.
Using the posterior on $\avec$, we construct the posterior distribution on the intrinsic pulsar noise and GW background power spectrum in each pulsar, which can now deviate from a power-law but are subject to a power-law prior distribution. 
We then test for deviations in the intrinsic pulsar noise spectrum for each pulsar and in the total GW background spectrum.

\subsection{Method}

\begin{figure*}
    \centering
    \includegraphics[width=\textwidth]{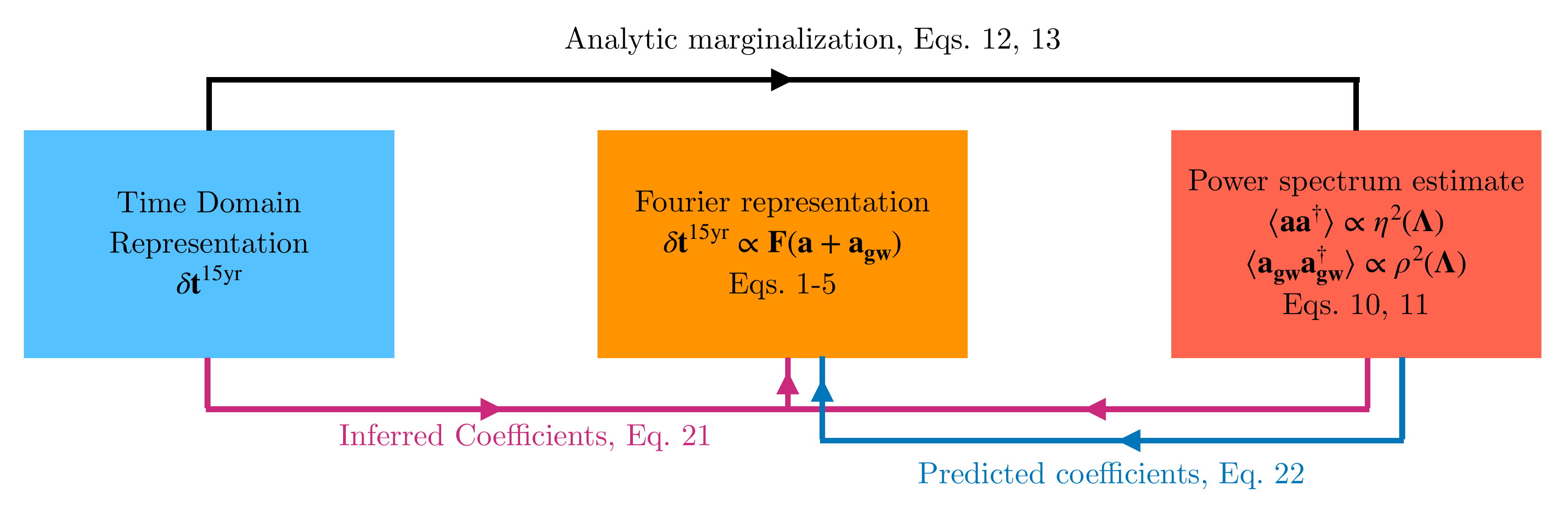}
    \caption{Workflow for the analysis of Sec.~\ref{sec:spectrum_tests} based on predicted and inferred Fourier coefficients for the GW and individual-pulsar red noise spectrum. 
    The black line corresponds to estimating $p(\bm\Lambda | \tresidng)$ directly after analytically marginalizing over the Fourier coefficients, and directly estimating the amplitude and spectral index for the power-law GW background and intrinsic pulsar noises. 
    The blue line corresponds to generating ``predicted'' coefficients using the power spectrum to simulate Fourier coefficients. 
    The maroon line indicates the inferred coefficients, which use the power-law power spectrum as a prior, combined with $\tresid$, to further constrain $\avec$.}
    \label{fig:flow_chart_mock_up}
\end{figure*}

In this subsection, we summarize the methods outlined in Ref.~\cite{ppc2}.
Given $p(\bm\Lambda|\tresidng)$, we calculate $\avec$ in two ways. 
In one method, $\avec$ are conditioned on $\tresidng$ and $\bm\Lambda$, which we refer to as the ``inferred'' coefficients (subscript ``inf'') because they are drawn from the inferred posterior on $\avec$ using information from both the power-law spectrum prior and the real data. 
In the other method, $\avec$ are conditioned only on $\bm\Lambda$, which we refer to as ``predicted'' coefficients (subscript ``pre'') because these are the coefficients predicted by the power-law spectrum prior. 
In both cases, we marginalize over $\bm\Lambda$ and $\epsvec$. 
We illustrate the workflow in Fig.~\ref{fig:flow_chart_mock_up}.
The posteriors on the inferred and predicted parameters are formally given by
\begin{align}
    p_{\rm inf}(\avec| \tresidng) &= \int d\bm\Lambda \, d\bm\epsilon\; p(\avec,\bm \epsilon | \bm\Lambda, \tresidng, \hd)\,\nonumber\\
    &\qquad\qquad\times p(\bm\Lambda| \tresidng,\hd)\,,
    \label{eq:bvec-observed}\\
        p_{\rm pre}(\avec| \tresidng) &= \int  d\bm\Lambda \, d\bm \epsilon\; p(\avec,\bm \epsilon | \bm\Lambda,\hd) p(\bm\Lambda | \tresidng,\hd)\nonumber\\
    &= \int  d\bm\Lambda \; p(\avec | \bm\Lambda, \hd) p(\bm\Lambda | \tresidng,\hd)\,.\label{eq:bvec-predicted}
\end{align}
The first term in the integrand differs between the predicted (no dependence on $\tresidng$) and inferred posteriors (which are conditioned on $\tresidng$). We discuss specifics of how these terms are evaluated below.
The second term in the integrand is the posterior on $\bm\Lambda$, e.g., power-law amplitudes and spectral indices.

In both Eqs.~\eqref{eq:bvec-observed} and~\eqref{eq:bvec-predicted}, we evaluate the posterior with a Monte Carlo integral.  
We first draw a sample (labeled with ``$s$'' superscript) $\bm\Lambda^s\sim p(\bm\Lambda | \tresidng)$.  
We then draw from the first term in the integrand. For the inferred coefficients we draw from $p(\avec, \epsvec | \bm\Lambda^s, \tresidng)$, which is a Gaussian with mean and covariance that depend on both the prior and the data, and are given by Eqs.~\eqref{eq:bvec_max_likelihood} and~\eqref{eq:bvec_covariance_matrix}. 
For the predicted coefficients, we draw from just the prior distribution (i.e., no dependence on data), $p(\avec,\epsvec | \bm\Lambda^s)$, which is a zero-mean Gaussian given by Eqs.~\eqref{eq:model_parameters_hyper_prior} and~\eqref{eq:big_b_matrix_definition}. 
More details on this scheme are discussed in Ref.~\cite{ppc2}.

For each $\bm\Lambda^s$ we split $\avec$ into $\avec_{\textrm{rn},a}^s$ for intrinsic pulsar noise, and $\avec_{\textrm{gw},a}^s$ for GWs for each pulsar $a$. 
We then reconstruct the power spectrum for the intrinsic pulsar noise in each pulsar and the GW power observed by each pulsar.
Each pulsar sees a different realisation of the GW background, but $\avec_{\textrm{gw},a}^s$ are drawn from the same distribution for all pulsars $a$. 
By contrast, the intrinsic pulsar noise is a different power spectrum for each pulsar, and therefore $\avec_{\textrm{rn},a}^s$ are drawn from a different distribution for each individual pulsar.

For the inferred coefficients, by conditioning on the data, the power spectrum will deviate from a power law if the true data-generating process differs from a power law. 
For the predicted spectrum, we obtain different realizations of a power spectrum that are consistent with a power-law. 
In Sec.~\ref{sec:individual_pulsar_spectra}, we discuss results for the inferred and predicted intrinsic pulsar noise and GW spectra for each pulsar, and compare them to an ``excess noise'' analysis done in Ref.~\cite{NG15detchar}. 
At each frequency, we use a modified version of the optimal statistic\footnote{See Sec.~IVB2 in Ref.~\cite{ppc2} for a discussion.} to combine individual-pulsar coefficients to estimate the total the total GW power across the PTA in that frequency bin~\cite{PFOS}. 
We present the results in Sec.~\ref{sec:gw_full_array_spectra}.

Finally, for each $\bm\Lambda^s$, we produce pulsar pair-wise correlations and compare them to the expected Hellings--Downs curve. 
To do this, we draw coefficients from $p_{\mathrm {inf}}(\avec|\tresidng, \hd)$ and $p_{\mathrm{pred}}(\avec|\tresidng,\hd)$, and use the optimal statistic to construct pair-wise correlations. 
In the case of the predicted parameters, this will give us an expected spread on correlation vs. angular separation for a given model.  
For the inferred parameters, by conditioning on the data, the correlations can deviate from the model. 
In Sec.~\ref{sec:full_array_spatial_correlations}, we look at reconstructions of the pair-wise correlations as a function of angular separation, and search for deviations from the Hellings--Downs curve.

\subsection{\label{sec:individual_pulsar_spectra}Power spectra of individual pulsars}

\begin{figure*}
    \centering
    \includegraphics[width=0.75\textwidth]{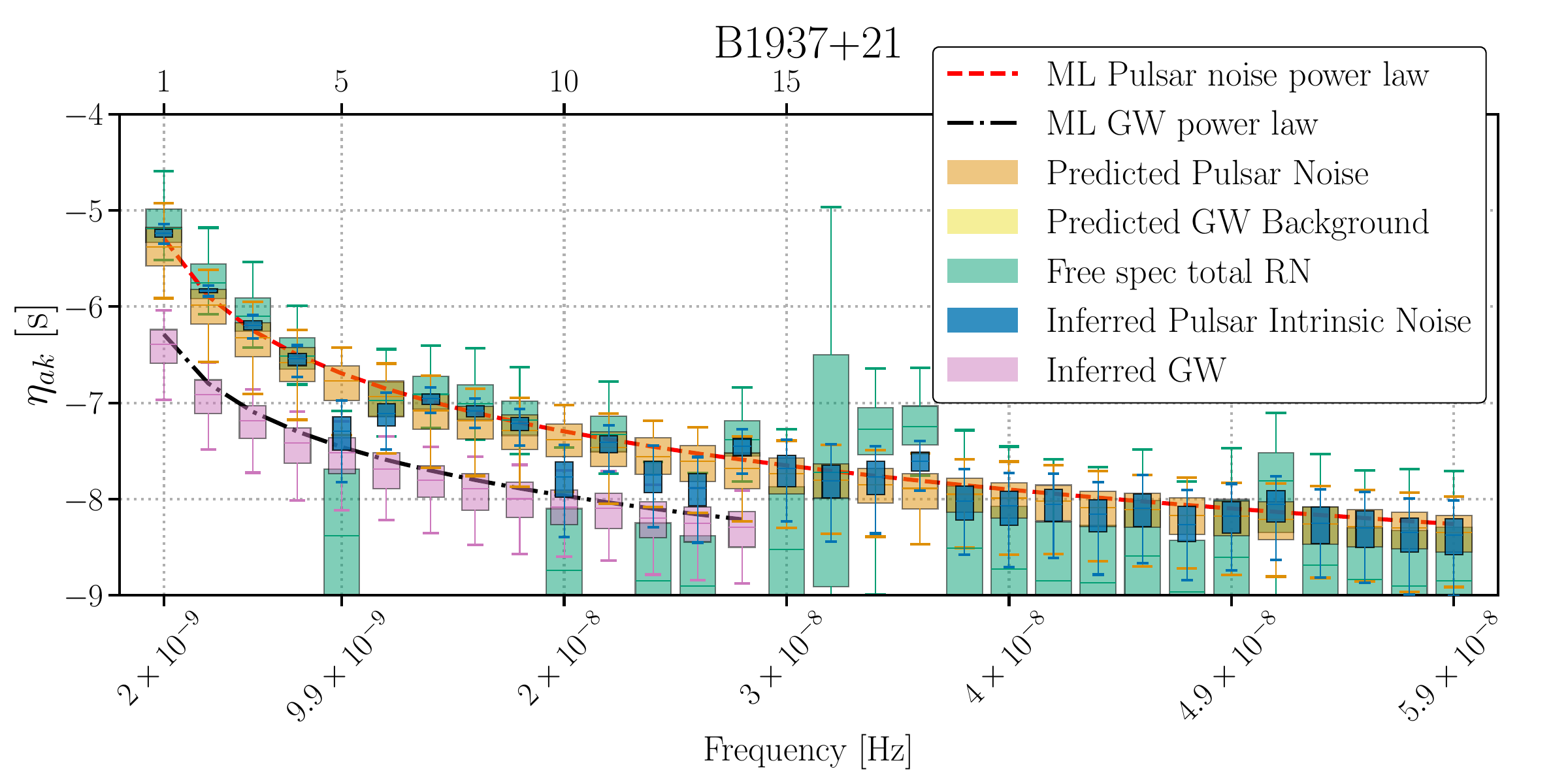}
    \includegraphics[width=0.75\textwidth]{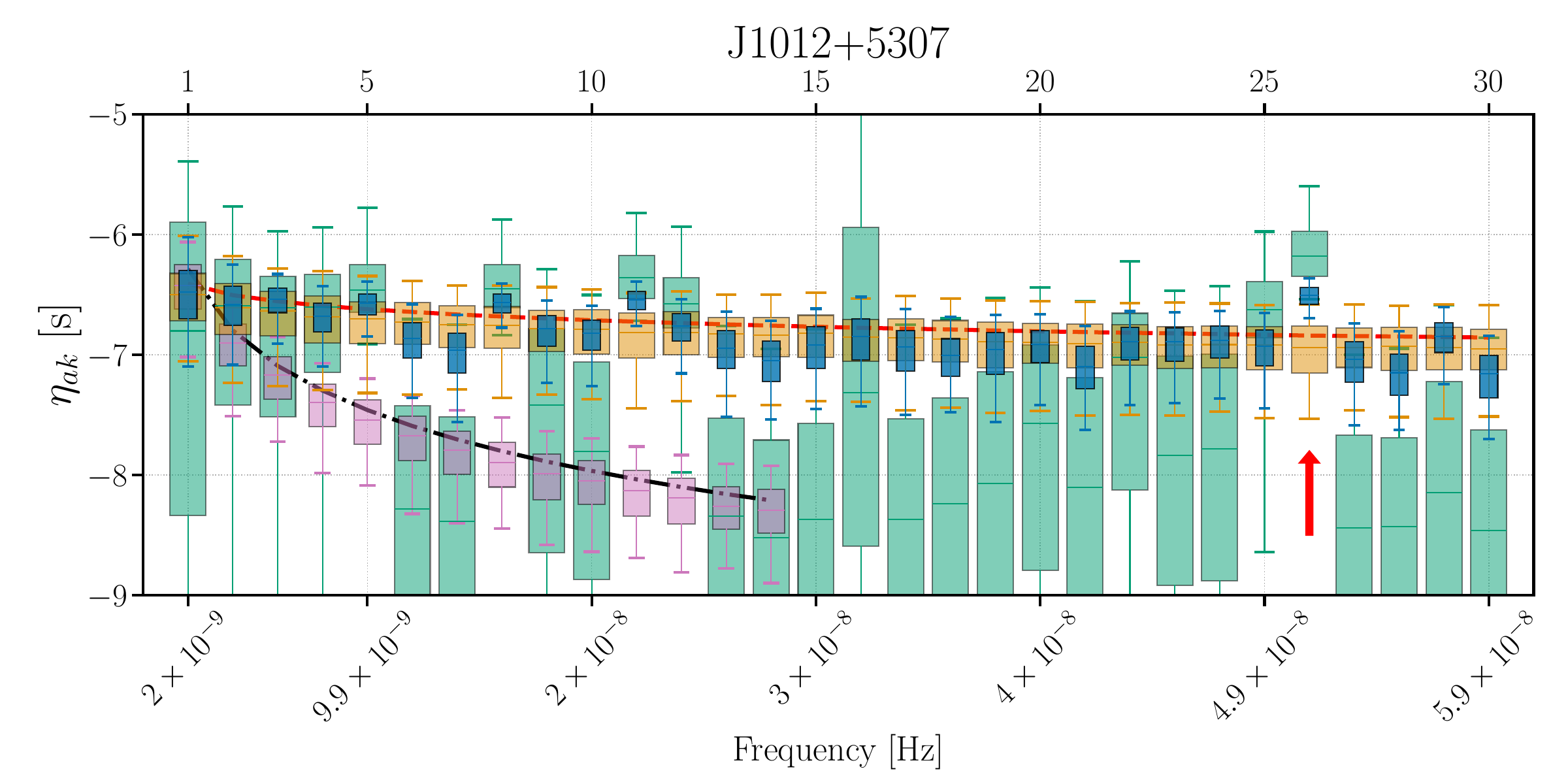}
    \includegraphics[width=0.75\textwidth]{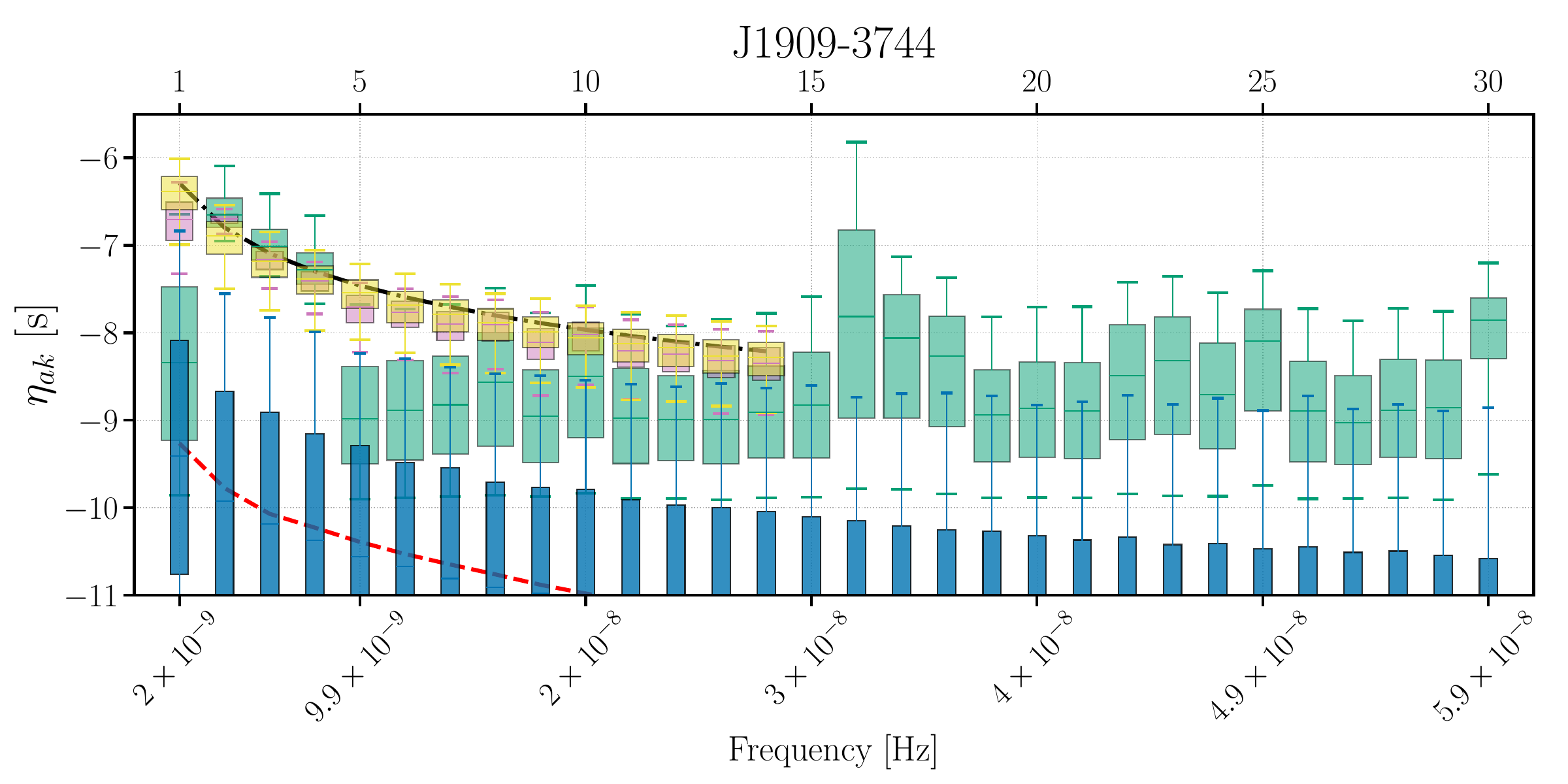}
    \caption{We show the total red noise (green, $\rho^2_i + \eta^2_{ai}$)~\cite{NG15detchar}, the inferred and predicted intrinsic pulsar noise (blue, $\eta^2_{ai,\textrm{inf}}$; orange, $\eta^2_{ai,\textrm{pred}}$) and inferred and predicted GW background (pink, $\rho^2_{ai,\textrm{inf}}$; yellow, $\rho^2_{ai,\textrm{pred}}$) for B1937+21 (top), J1012+5307 (middle), J1909-3744 (bottom). The boxes indicate the 50\% credible interval, while the whiskers show the 5th and 95th percentiles. The red dashed line shows the maximum likelihood (ML) intrinsic pulsar noise power-law, and the black dashed line shows the same for the GW background. To reduce clutter, we only show $\eta^2_{ai,\textrm{pred}}$ in the top two panels, and $\rho^2_{ai,\textrm{pred}}$ in the bottom panel. The top two pulsars exhibit strong intrinsic pulsar noise, that is larger than the estimated background, because the greeen and blue boxes are larger than the pink ones. In the bottom panel, total red noise (green) is dominated by the GW background (pink, yellow), while the intrinsic pulsar noise (blue) is not detectable.  In some frequency bins, there appears to be a lack of red noise, but this is consistent with what is expected from a power-law model. There is little evidence for excess noise, the strongest evidence being the $26^{\textrm{th}}$ bin for J1012+5307 (marked with a red arrow), which has a Bayesian $p$-value of 0.03, which we discuss in the text. }
    \label{fig:rn_comparison}
\end{figure*}

\begin{figure*}
    \centering
    \includegraphics[width=0.9\textwidth]{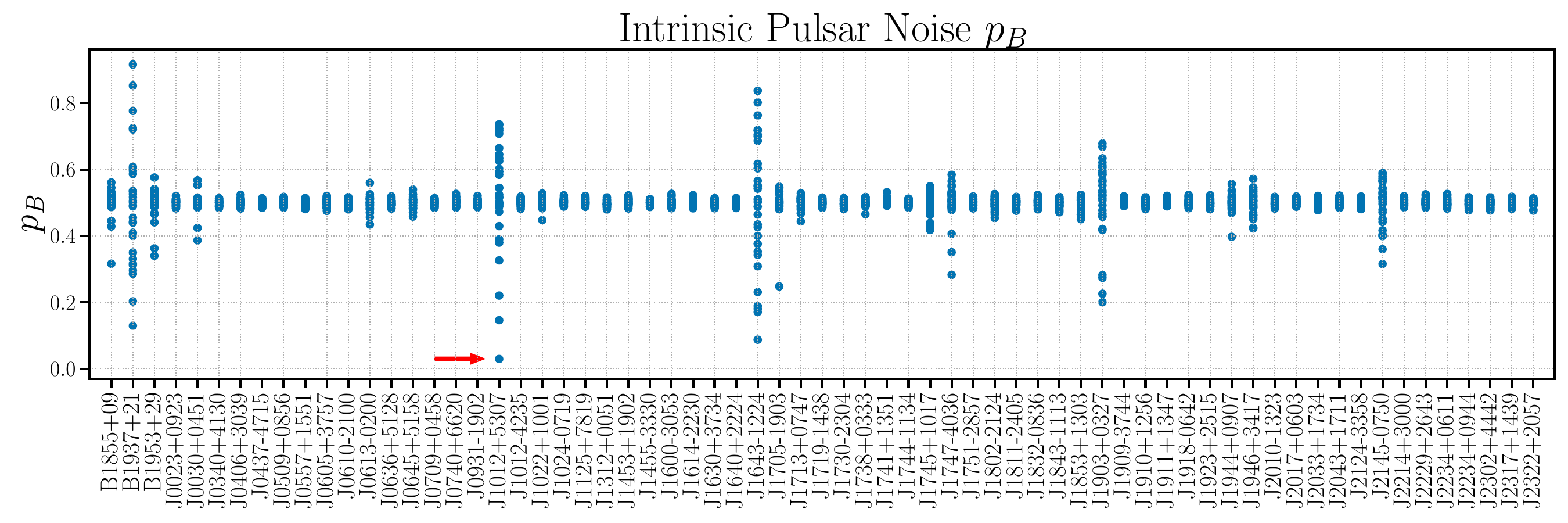}
    \includegraphics[width=0.9\textwidth]{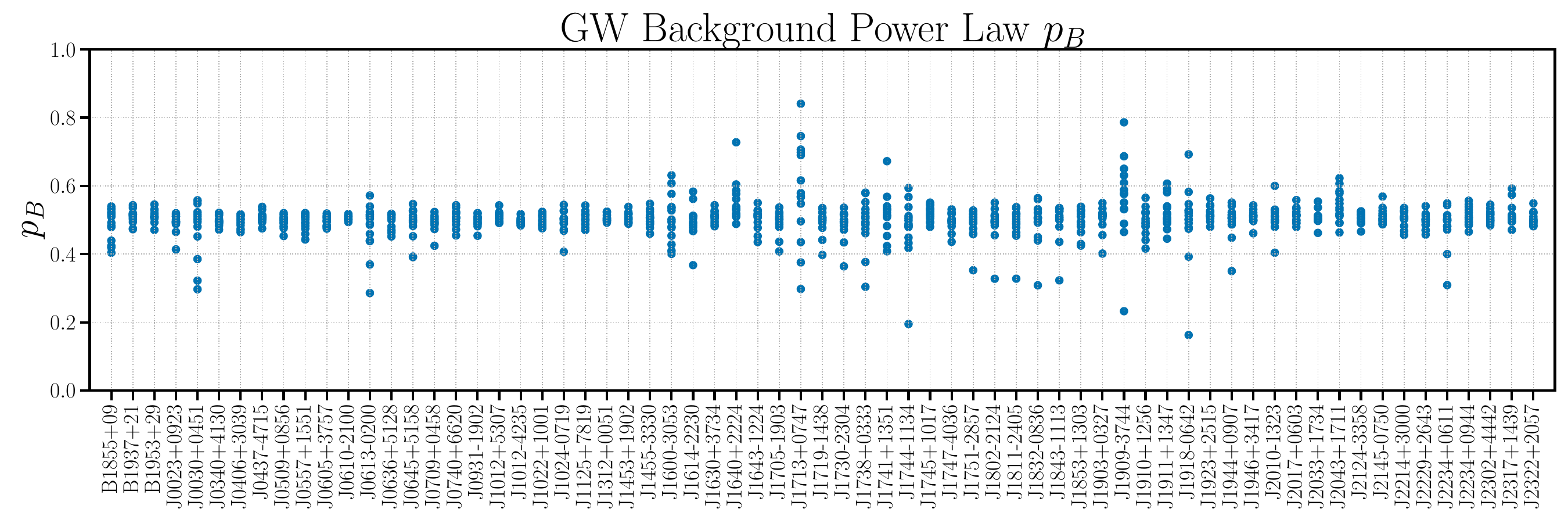}
    \caption{For each pulsar on the horizontal axis we show $p_B$ for all 30 (14) frequency bins in the top (bottom) panel. For most pulsars $p_B$ is near 0.5, indicating that the inferred and predicted power spectra agree with one another, and so a power-law is an appropriate model for the intrinsic pulsar noise. For several pulsars, there is a broader spread in $p_B$, including two of the pulsars that we show in Fig.~\ref{fig:rn_comparison}}.
    \label{fig:pb_for_ipn_each_pulsar}
\end{figure*}

Power spectra for each pulsar in the NANOGrav array are explored in Ref.~\cite{NG15detchar}, it is therefore worth contrasting the two results.
First, Ref.~\cite{NG15detchar} simultaneously estimated the \textit{total} red noise ($\rho_i^2 + \eta_{ai}^2$) in each frequency bin $i$, performing a separate analysis for each individual pulsar $a$. 
A Savage-Dickey Bayes factor was calculated to estimate the significance of the total red noise at each frequency for each pulsar. 
Next, both the common red noise and intrinsic pulsar noise were fixed to the maximum likelihood values estimated from a \curn analysis that assumes these spectra follow a power-law. 
Once fixing these parameters in their model, \textit{excess noise} in each frequency bin for each pulsar was searched for. No evidence for excess noise was found, the power-law model for intrinsic pulsar noise and the common red noise processes are therefore sufficient.

In this work, we instead separate intrinsic pulsar noise and GW background contributions when drawing parameters from the inferred and predicted distributions, and we produce a posterior distribution on both contributions in each frequency bin for each pulsar. 
This way we are testing both the intrinsic pulsar noise and GW background power-law assumptions at the same time, while constructing full posteriors on the intrinsic pulsar noise and the GW background. 
This is in contrast to the search for excess noise on top of a power-law common red noise and intrinsic pulsar noise. 
Another difference is that in this work, individual frequency bin estimates are subject to a prior that follows a power-law, while Ref.~\cite{NG15detchar} used a log-uniform prior on the power in each frequency bin.

In Fig.~\ref{fig:rn_comparison}, we show results for two pulsars with strong intrinsic pulsar noise, B1937+21 and J1012+5307 (top two panels), and J1909-3744 which has no measurable intrinsic pulsar noise, but a contribution attributed to the GW background.
The green boxes correspond to the estimates of the total red noise power from~\cite{NG15detchar}, which was discussed above. 
The blue and pink boxes correspond to the inferred intrinsic pulsar noise and GW background contributions respectively, orange boxes correspond to the predicted intrinsic pulsar noise, and yellow boxes correspond to the predicted GW background in the bottom panel. Similar plots for each pulsar are included in an electronic supplement.

In the top two panels, the intrinsic pulsar noise (inferred in blue boxes, predicted in orange boxes) typically agrees with the total red noise (green boxes) from Ref.~\cite{NG15detchar}--indicating that the total red noise is dominated by intrinsic pulsar noise. 
The GW background is significantly below the intrinsic pulsar noise and the total red noise. 
Additionally, the orange distributions, corresponding to predicted intrinsic pulsar noise, agree with the inferred intrinsic pulsar noise. 
We quantify this agreement below.
In the bottom panel, the total red noise agrees with the GW background, while the intrinsic pulsar noise is significantly lower. 
This is consistent with the GW background contributing significantly to $\tresidng$ in this pulsar, with no intrinsic pulsar noise contribution. 
In a few cases the free spectrum total red noise (green boxes) deviates further from the power law than the inferred intrinsic pulsar noise (blue boxes). This is due to the power-law prior used for the inferred intrinsic pulsar noise, which will tend to move those parameters closer to the power law.

In a few situations, e.g., the first and fifth through seventh bins for J1909-3744, the estimated total red noise (green boxes) appears to be lower than the predicted (yellow boxes) and inferred spectra (pink boxes) for the GW background. The low total noise are consistent with the predicted distributions, which follow a $\chi^2$ with two degrees of freedom, and therefore have large support at those low values (despite what the box and whiskers show). 
The inferred distributions broadly agree with the predicted, and do show overlap with the green whiskers; but they do look inflated compared to the total noise. 
However, simply combining the GW background and intrinsic pulsar noise spectra in each bin may not reproduce the green boxes for a few reasons. 
First, interference between these two contributions may cancel or amplify the estimated total red noise above or below what we would expect from naively adding their contributions incoherently. 
Second, a log-uniform prior on the power in each frequency will likely reduce the upper limit on the estimated power in that bin when no red noise detection is made when compared to the upper limit we would set using a prior informed by the power law model

We then quantify excess intrinsic pulsar noise in individual frequency bins for each pulsar. For each pulsar $a$ and frequency bin $i$ we calculate
\begin{align}
p_B = \frac{1}{N}\sum_{s=1}^{N} \Theta(\eta^2_{\textrm{inf},ai}-\eta_{\textrm{pred},ai}^2)\,,
\end{align}
where $\eta^2_{\textrm{inf},ai}$ are drawn from $p_{\rm inf}(\avec| \tresidng)$, and correspond to the inferred power spectrum, while $\eta_{\textrm{pred},ai}^2$ are drawn from $p_{\rm pred}(\avec| \tresidng)$ and correspond to the predicted power spectrum due to a power law. 
Each $\eta^2$ carries with it an implicit $s$ index, which we have suppressed.
We also use the \hdposterior simulations to calculate
\begin{align}
p_B^{\mathrm{sim}} = \frac{1}{N}\sum_{s=1}^{N} \Theta[\eta^2_{\textrm{inf},ai}-(\eta_{\textrm{inf},ai}^2)^\mathrm{rep}]\,,
\end{align}
where the superscript ``rep'' indicates it is the inferred estimate on the power in that frequency calculated on the replicated data. Note that $\eta^2_{\textrm{inf},ai}$ and $(\eta_{\textrm{inf},ai}^2)^\mathrm{rep}$ are calculated using the same $\bm\Lambda^s$. We find that these two methods produce nearly identical results, and so we report results for $p_B$ instead of $p_B^{\rm sim}$.

For intrinsic pulsar noise across all pulsars and frequencies, we find a minimum of $p_B=0.03$, for J1012+5307, $f=51\,\textrm{nHz}$, which is the box that is visibly above the max likelihood curve in the middle panel of Fig.~\ref{fig:rn_comparison}, marked with the red arrow. 
This is not a significant $p$-value, given that we are analyzing 67 pulsars and 30 frequency bins, and so we cannot conclude that this represents a deviation from a power law. This is the same conclusion as Ref.~\cite{NG15detchar}. The minimum and maximum $p_B$ for the GW background power spectrum across all pulsars are 0.16 and 0.84. 

Deviations from the power-law model may not just take the form of excess noise at individual frequencies. For example, one could have a broken power-law, or excess noise across multiple frequencies that are not individually detectable. 
We do not develop a statistic to measure this here, as it requires a specific model to compare to the power-law model and there are a broad range of potential models. 
However, such an analysis should be done in the future. 

We show a plot of $p_B$ for intrinsic pulsar noise for each pulsar in the top panel of Fig.~\ref{fig:pb_for_ipn_each_pulsar} and for the GW background in each pulsar in the bottom panel. 
The horizontal axis corresponds to each pulsar, while the vertical axis corresponds to $p_B$; each point represents a $p_B$ for each pulsar and each frequency bin. 
When the inferred and predicted power spectrum estimates agree at a given frequency bin, we expect $p_B\approx 0.5$. 
We see that a few noisy pulsars show $p_B$ values that stray away from 0.5, including the pulsars in the top two panels of Fig.~\ref{fig:rn_comparison}. 
As stated before, no individual frequency bin shows an extreme value of $p_B$, e.g., $p_B < 0.01$ or $p_B > 0.99$, meaning we cannot state there are individual bins with excess noise. 
However, pulsars like B1937+21 and J1012+5307 do show quite a few frequency bins with $p_B$ deviating from 0.5, meaning they may benefit from a more flexible model in the future. 
It is currently prohibitively computationally expensive to estimate intrinsic pulsar noise separately in each frequency bin for each pulsar when we estimate $p(\bm\Lambda | \tresidng)$. 
However, with future computational improvements, we may be able to do this for a limited number of pulsars, and this method provides a good starting point for choosing those pulsars. 


\subsection{\label{sec:GW}Full-Array Results for the Gravitational-wave Background}

\begin{figure}
    \centering
    \includegraphics[width=\columnwidth]{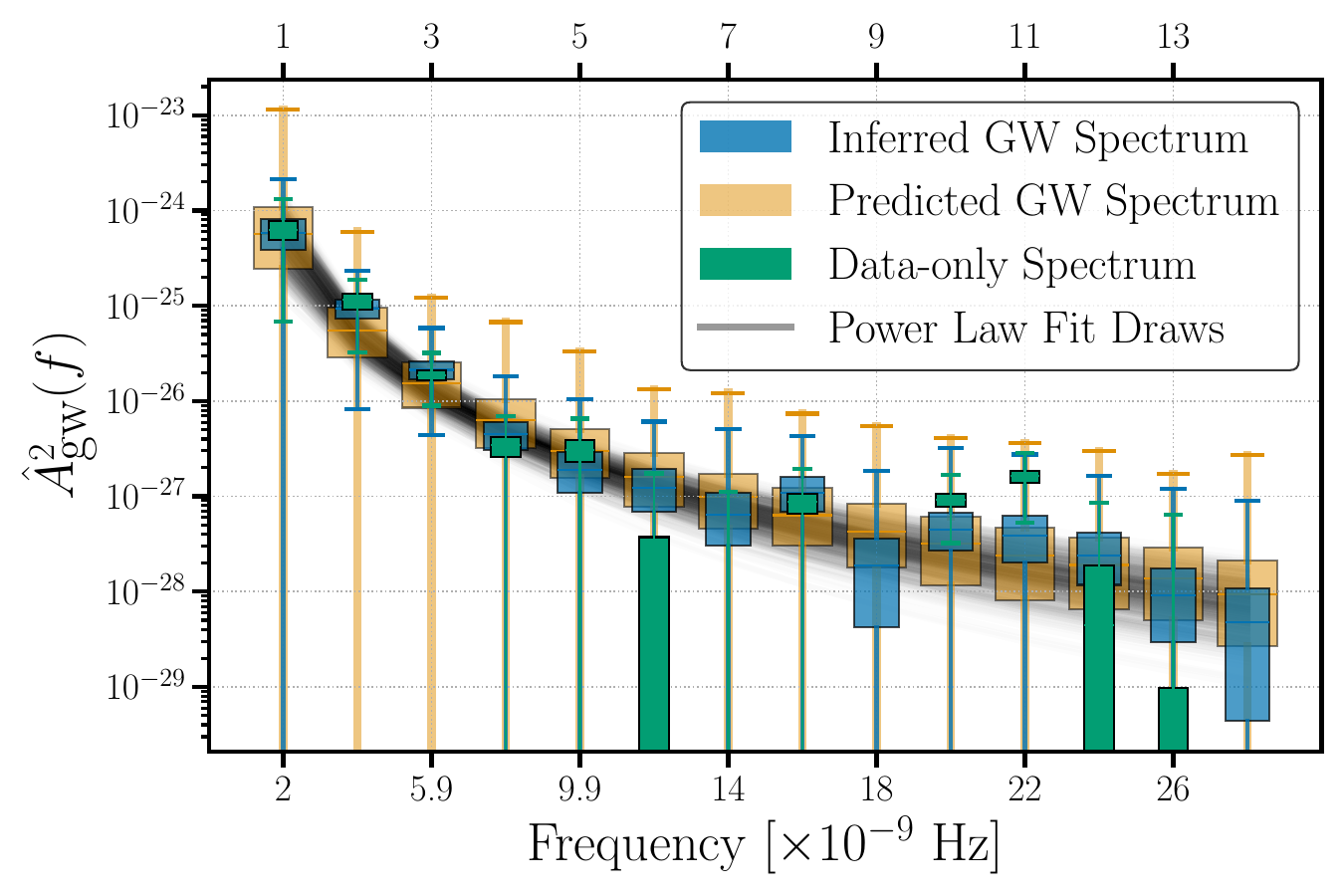}
    \caption{We show reconstructed GW power in each frequency bin for inferred (blue) and predicted (orange) coefficients, and a data-only (green) reconstruction. The boxes correspond to inter-quartile ranges and the whiskers are the 5th and 95th percentiles. Power-law draws from $p(\bm\Lambda | \tresidng)$ are shown in light black. The blue and orange distributions agree with one another at most frequencies. In a few places, the green boxes differ from the predicted or inferred distribution (e.g., 20 and 22\,nHz), but the data are weakly informative, as the inferred and predicted distributions agree with one another in those cases. In a few frequencies the data only distribution shows evidence for negative power, this is to be expected when the data are not informative (discussed further in the text).}
    \label{fig:all_psr_spectrum_estimate}
\end{figure}

\subsubsection{Spectral shape}
\label{sec:gw_full_array_spectra}
To get a full-PTA estimate of the GW background power in each frequency bin, we use a modified version of the optimal statistic~\cite{chamberlinOS,Vigeland:2018ipb,ppc2} that estimates the Hellings--Downs correlated GW power in each individual frequency bin. The details of this statistic are discussed in Sec. IV B 2 of Refs.~\cite{ppc2} and~\cite{PFOS}.

In Fig.~\ref{fig:all_psr_spectrum_estimate} we show results for the estimated power in each frequency bin for the inferred parameters (blue), the predicted parameters (orange), and a fully frequentist estimate that depends only on the data (green). 
The boxes correspond to the interquartile range of the GW power in each bin, estimated over draws from $p(\bm\Lambda| \tresidng)$, and the whiskers are the 5th and 95th percentiles.
In showing these together, we compare predicted, inferred, and data-only results. 
There is no visible evidence for a deviation from a power law, which is indicated by the gray shaded region, which encompasses draws from $p(\bm\Lambda| \tresidng)$. 
The inter-quartile ranges for the predicted, inferred, and data-only power overlap in most frequency bins.
In a few places, the data-only results appear to differ from the inferred and predicted results, e.g., 9th--11th bins, but the data are weakly informative and the inferred parameters are closer to the predicted parameters.

The per-frequency optimal statistic, used to combine $\avec_{\textrm{gw}}$ across pulsars allows for negative power in situations where the data are uninformative. Like the traditional optimal statistic, when no correlated signal is present, the distribution of the statistic is GX2 centered at zero.
This is why the whiskers for several frequencies leave the bottom of the plot, and in two cases (bins 9 and 14) the data-only inter-quartile ranges are negative. This does not change our conclusions, as we find that our results at frequencies where we know we should see correlated GW power show such power (specifically the lowest five frequency bins). This is consistent with the \hd free-spectrum results in Ref.~\cite{NANOGrav:2023gor}.

We use the \hdposterior data replications to compare simulations from a power-law model to the results in Fig.~\ref{fig:all_psr_spectrum_estimate}. 
We find that the spectral results are consistent with a power-law model with Hellings--Downs correlations--the lowest and highest $p_B$ comparing inferred parameters from simulations with inferred parameters on the 15 yr data set are 0.30 and 0.72.

\subsubsection{Spatial Correlations}
\label{sec:full_array_spatial_correlations}

\begin{figure}
    \centering
    \includegraphics[width=\columnwidth]{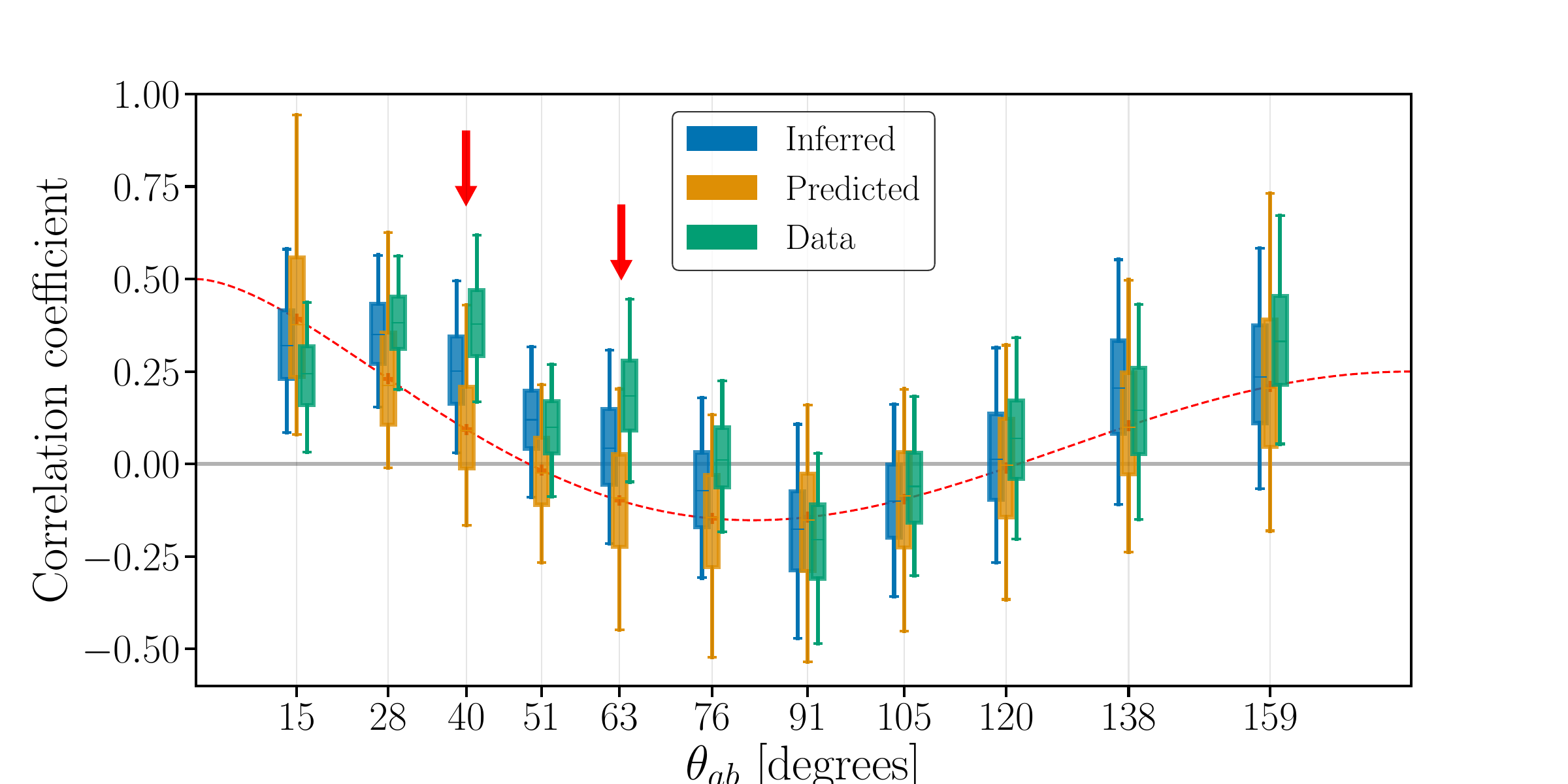}
    \caption{We show reconstructed binned spatial correlations. The predicted (orange) show the expected spread around the Hellings--Downs curve that we might expect across many realizations. The inferred (blue) and data-only (green) recoveries broadly agree with the orange. There are two bins (3rd and 5th) that show some deviation between the green and orange bins. We find that these bins are not statistically significant when comparing the inferred and predicted distributions, and that changing the binning does not have a significant affect.}
    \label{fig:hd_reconstruction_hd_prior}
\end{figure}

In Sec.~\ref{sec:bayes_pval_os}, we showed broad consistency between the data and the \hd model, and we showed that there is no evidence for additional monopolar or dipolar correlations. 
Those tests compare plausible alternative analytic correlation models to the expected correlation model. 
In this section, we search for isolated deviations in the binned spatial correlations from the Hellings--Downs curve. 
We use the optimal statistic on the inferred and predicted coefficients, as well as directly on the data, and compare the inferred, predicted, and data-only binned reconstructions to search for potential deviations from the Hellings--Downs curve that are not just monopolar or dipolar.

To construct the binned correlations, we perform an inverse--noise-weighted average over correlations for all pulsar pairs whose angular separation falls in a given angular-separation bin. 
An example for 11 bins of equal width is shown in Fig.~\ref{fig:hd_reconstruction_hd_prior}. 
We compare the inferred (blue), predicted (orange), and data-only (green) estimates of the binned correlations. 
The spread comes from calculating the mean and variance of the correlation across pulsar pairs for a given posterior draw, and sampling from a univariate Gaussian with that mean and variance, and then repeating over many draws from $p(\bm\Lambda | \tresid).$ 
The variance for the bin for a given draw includes covariance between pairs of pulsars due to the non-zero GW background~\cite{Allen:2022ksj}. 
The bars indicate the 5th and 95th percentiles of the resulting distribution.

We find that the data and inferred correlations are consistent with the predicted correlations. 
We also use the \hdposterior replications and find that none of the inferred or data-only binned correlations differ significantly from those calculated with the data replications. 
The two bins with the most extreme $p_B$\footnote{Because we are comparing two distributions, we consider both large and small $p$-values to be extreme.} are the third and fifth bins, with $p_B=0.87,\, 0.88$ respectively, indicating that, if we take Hellings--Downs correlations as our prior, we do not have evidence for deviations from the Hellings--Downs curve.
This does not mean that we are fully consistent with Hellings--Downs correlations (as subtle changes in each bin could result in a different overall correlation pattern), but it does indicate that there are no obvious ``spikes'' in correlations on small angular scales. Changing the choice of binning does not change the qualitative conclusion.


\section{\label{sec:leave_one_out}Leave-one-out Analyses}

\begin{figure*}
    \centering
    \includegraphics[width=\textwidth]{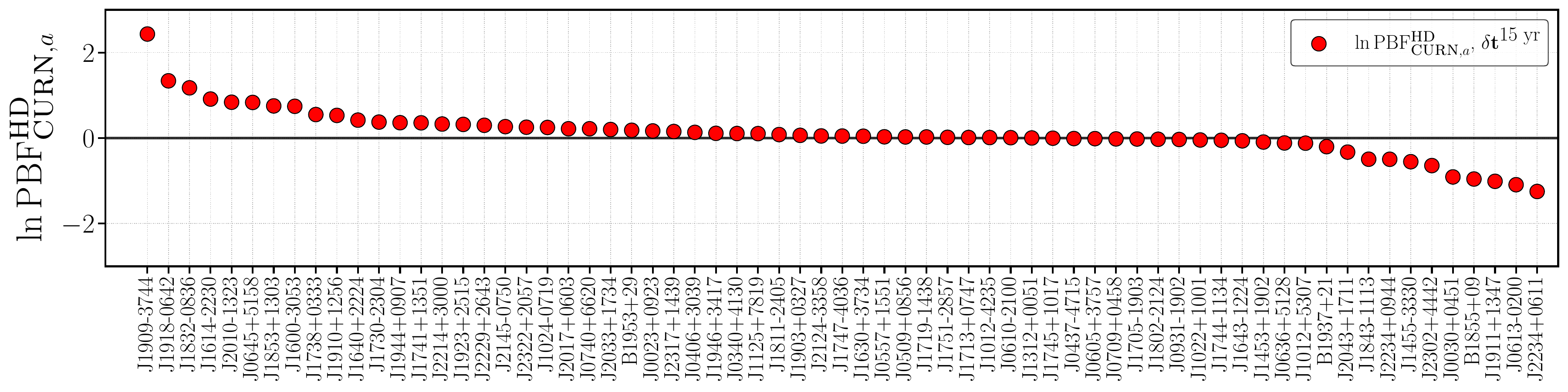}
    \caption{We show $\ln\textrm{PBF}_{\textrm{\curn},a}^{\textrm{\hd}}$ the 67 pulsars. The pulsar best predicted by the \hd model is J1909$-$3744, whose red noise is predominantly due to a GW background. There are 43 pulsars with $\ln\textrm{PBF}_{\textrm{\curn},a}^{\textrm{\hd}}>0$ and 25 with $\ln\textrm{PBF}_{\textrm{\curn},a}^{\textrm{\hd}} < 0.$ As discussed in the text, we find the number and level of the pulsars with $\ln\textrm{PBF}_{\textrm{\curn},a}^{\textrm{\hd}}<0$ to be consistent with simulations that have a GW background as strong as what we find in the data.}
    \label{fig:dropout_results}
\end{figure*}

Although individual pulsars can exhibit unique chromatic noise features, profile changes, and red noise properties, similar noise models are fit to each pulsar in the array. 
In this section, we seek to identify whether certain pulsars are poorly fit by these models.
We perform a leave-one-out analysis, where we calculate a posterior predictive likelihood for the timing residuals in one pulsar, given the data in all other pulsars~\cite{ppc2}. This analysis is similar to the one in Ref.~\cite{NANOGrav:2023gor}, with a few key differences. 
First, we use 14 frequency bins for the analysis, and we include the negative spectral index of the GW background, $\gamma$, in the initial fit.  In Ref.~\cite{NANOGrav:2023gor}, $\gamma$ is fixed to $13/3$. We also use a larger number of \curn simulations to evaluate the significance of the GW background, and perform a new comparison between simulated and real data on each individual pulsar. 

Both the \curn and the \hd models can be used to predict features in one pulsar, given a model fit to the other pulsars in the array. 
The \curn model, for example, can only predict the variance of the common-process--induced timing residuals, while the \hd model, which includes GW-induced correlations, makes a prediction for both the variance of the timing residuals and their waveform.\footnote{This prediction is limited by the strength of the Hellings--Downs correlations. We cannot predict the pulsar-term fluctuations, but Earth-term predictions are informative.}

We compare the predictive power of these two models by calculating a pseudo Bayes factor (PBF), which is the ratio of the posterior predictive likelihood for the \hd and the \curn models. 
We calculate this on both a pulsar-by-pulsar basis, to identify potential pulsars that are not well-predicted by the models, and also across the full pulsar timing array to construct a new detection statistic. 

We denote the ``left-out'' pulsar with subscript $a$ and the rest of the data set excluding that pulsar with a subscript $-a$. The posterior predictive likelihood is

\begin{align}
\label{eq:posterior_predictive_likelihood}
    p(\tresid_a | \tresid_{-a}) = \int \dd\bm\Lambda \,\dd\avec\,\dd\epsvec\;p(\tresid_a | \avec, \epsvec, \bm\Lambda)p(\avec, \epsvec, \bm\Lambda | \tresid_{-a})\,.
\end{align}
As in Ref.~\cite{ppc2}, we split up the parameters and hyperparameters into separate pieces based on whether they correspond to pulsar $a$ or pulsars $-a$, and whether they describe GW or red noise coefficients, $\bm\Lambda = [\bm\Lambda_a, \bm\Lambda_{-a}, \bm\Lambda_{\textrm{gw}}]$, $\avec = [\avec_{\textrm{gw},a}, \avec_{\textrm{gw},-a}, \avec_{a}, \avec_{-a}]$. We use this new notation, and evaluate Eq.~\eqref{eq:posterior_predictive_likelihood} for the \hd and the \curn models to find, see App. A of~\cite{ppc2}, 
\begin{align}
    \label{eq:hd_ppl}
    \nonumber p_\textrm{\hd}(\tresid_a | \tresid_{-a}) \approx \frac{1}{N_s} \sum_{s=1}^{N_s}\int &\dd\bm\Lambda_a \dd \avec_{\textrm{gw},a} \; p(\tresid_a | \bm\Lambda_a, \avec_{\textrm{gw}, a})\\
    \nonumber &\times p(\avec_{\textrm{gw}, a} | \bm\Lambda_{\textrm{gw}}^{s},\bm\Lambda_{-a}^s,\tresid_{-a})\\
    & \times p(\bm\Lambda_{a})\,,\\
    \label{eq:curn_ppl}
    p_\textrm{\curn}(\tresid_a | \tresid_{-a}) \approx \frac{1}{N_s} \sum_{s=1}^{N_s}\int &\dd\bm\Lambda_a p(\tresid_a | \bm\Lambda_a, \bm\Lambda_{\textrm{gw}}^s)p(\bm\Lambda_a)\,.
\end{align}
In both cases, we perform a Monte Carlo integral over the hyperparameter posterior
\begin{align}
    \bm\Lambda_{\textrm{gw}}^s,\bm\Lambda_{-a}^s \sim p(\bm\Lambda_{\textrm{gw}}^s,\bm\Lambda_{-a}^s | \tresid_{-a})\,.
\end{align}
The main difference between Eq.~\eqref{eq:hd_ppl} and Eq.~\eqref{eq:curn_ppl} is that for the $\hd$ model, the $-a$ pulsars can produce a prediction for $\avec_{\textrm{gw}, a}$ due to the Hellings and Downs correlations, while the \curn model cannot.

The ratio of the posterior predictive likelihoods for the \curn and \hd models, is the PBF and it can be used to compare the two models. We first calculate the PBF point-wise across pulsars
\begin{align}
    \textrm{PBF}_{\textrm{\curn},a}^{\textrm{\hd}} = \frac{p_\textrm{\hd}(\tresid_a | \tresid_{-a})}{p_\textrm{\curn}(\tresid_a | \tresid_{-a})}\,,
\end{align}
and then the \textit{total} PBF as a point-wise product
\begin{align}
    \textrm{PBF}_{\textrm{\curn}}^{\textrm{\hd}} = \prod_a \textrm{PBF}_{\textrm{\curn},a}^{\textrm{\hd}}\,.
\end{align}
A full discussion of the differences and similarities between a typical Bayes Factor and the PBF is given in Ref.~\cite{ppc2}, but we summarize a few key points here. 
Unlike the Bayes Factor, the PBF is not sensitive to parts of the parameter space that have no likelihood support. 
The PBF compares how well the models predict new data, while the Bayes factor is a summary statistic comparing how well two models fit existing data. 
Both statistics, however, are uncalibrated--meaning it is unclear how to interpret statistical significance as a function of the value of the statistic. 
In this section, similar to previous sections, we use data replications to assess the significance of the PBF. 

Importantly, we can calculate the PBF on each pulsar individually, and identify whether certain pulsars are ``outliers'' that are not well-predicted by a given model. This is similar to the ``dropout factor'' analysis in~\cite{NANOGrav:2020bcs,NANOGrav:2023gor}. In this work, we calculate a separate predictive likelihood for each model for each pulsar, while the dropout factor analysis samples an indicator variable that chooses whether to model a pulsar with the \curn or \hd model. The interpretation of the results are similar to point-wise results.

Across the array, multiplying all of the ``leave-out" PBFs we find $\textrm{PBF}_{\textrm{\curn}}^{\textrm{\hd}}=\pbfFullArray$.
This is on a similar scale to the 14 frequency Bayes factor comparing \hd and \curn~\cite{NANOGrav:2023gor}, but as with typical Bayes factors, there is no natural scale to use to ``calibrate'' this level of significance. 
In Ref.~\cite{NANOGrav:2023gor}, several methods are used to generate a null distribution for detection statistics, including sky scrambles~\cite{Cornish:2015ikx}, phase shifts~\cite{Taylor:2016gpq}, and simulated data sets.
In this work, we again resort to simulated data sets. 
Using 600 \textsc{CURNPosteriorDraws} simulations, we calculate $\textrm{PBF}_{\textrm{\curn}}^{\textrm{\hd}}$ on each of the simulations. 
We find a Gaussian equivalent $p$-value of $\pbfSimulationSigmaVal\,\sigma$ in favor of Hellings--Downs correlations on the 15 yr NANOGrav data. 

We also use the \hdposterior draws to test whether this result is consistent with the \hd model.
We find that $\textrm{PBF}_{\textrm{\curn}}^{\textrm{\hd}}$ falls in the 27th percentile of the \hd simulations, again confirming that our results are inconsistent with the \curn model, and are consistent with the \hd model.

We show $\ln \textrm{PBF}_{\textrm{\curn},a}^{\textrm{\hd}}$ for each pulsar in Fig.~\ref{fig:dropout_results}. 
There are more pulsars with $\ln \textrm{PBF}_{\textrm{\curn},a}^{\textrm{\hd}}>0$ than the reverse, because the \hd model is better at predicting new data than the \curn model.
There are several pulsars with $\ln \textrm{PBF}_{\textrm{\curn},a}^{\textrm{\hd}}<0$. We expect this in a few pulsars due to the specific realization of intrinsic pulsar noise and the pulsar term from the GW background, which we cannot predict.
To understand whether the number of pulsars with $\ln \textrm{PBF}_{\textrm{\curn},a}^{\textrm{\hd}}<0$ is expected, and whether the typical scale of those downward fluctuations is ``representative'' of what we would expect from a GW background, we perform simulations.  
We do 200 \hdposterior simulations and calculate $\ln \textrm{PBF}_{\textrm{\curn},a}^{\textrm{\hd}}$ for each of those simulations to understand what the typical PBF is for the ``best'' and ``worst'' predicted pulsars if we have a GW background consistent with our posteriors.

\begin{figure*}
    \centering
    \includegraphics[width=\textwidth]{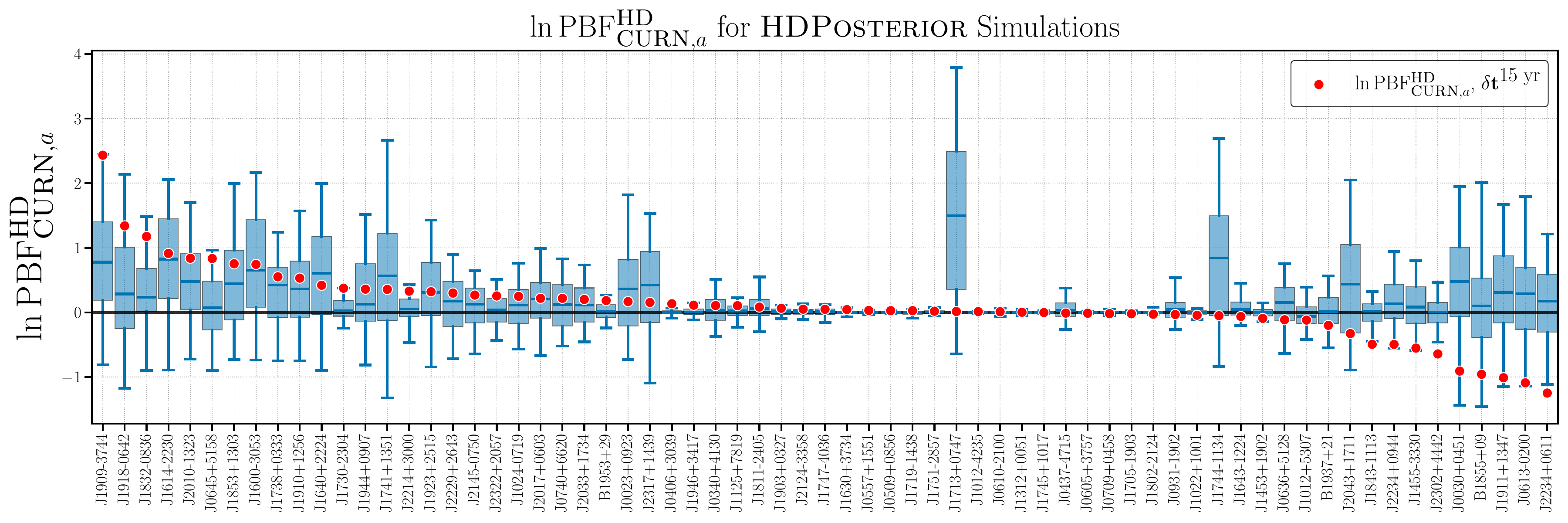}
    \caption{We show $\ln\,\textrm{PBF}_{\textrm{\curn},a}^{\textrm{\hd}}$ on $\tresidng$ in red. The blue box and whisker plot shows the interquartile range and 5th and 95th percentiles of the distribution of $\ln\,\textrm{PBF}_{\textrm{\curn},a}^{\textrm{\hd}}$ for 200 of the \hdposterior simulations for each pulsar. For most pulsars the median falls above zero for these simulations, indicating that $\hd$ is the preferred model, as expected. Calculating the percentile of the red point within the blue distribution for each pulsar yields a set of percentiles that are consistent with a uniform distribution between 0 and 1, which means $\textrm{PBF}_{\textrm{\curn},a}^{\textrm{\hd}}$ on $\tresidng$ is consistent with what we expect from a model with \hd correlations and intrinsic pulsar noise consistent with $p(\bm\Lambda | \tresidng)$.}
    \label{fig:pbf_simulation_results}
\end{figure*}
\begin{figure*}
    \centering
    \includegraphics[width=\textwidth]{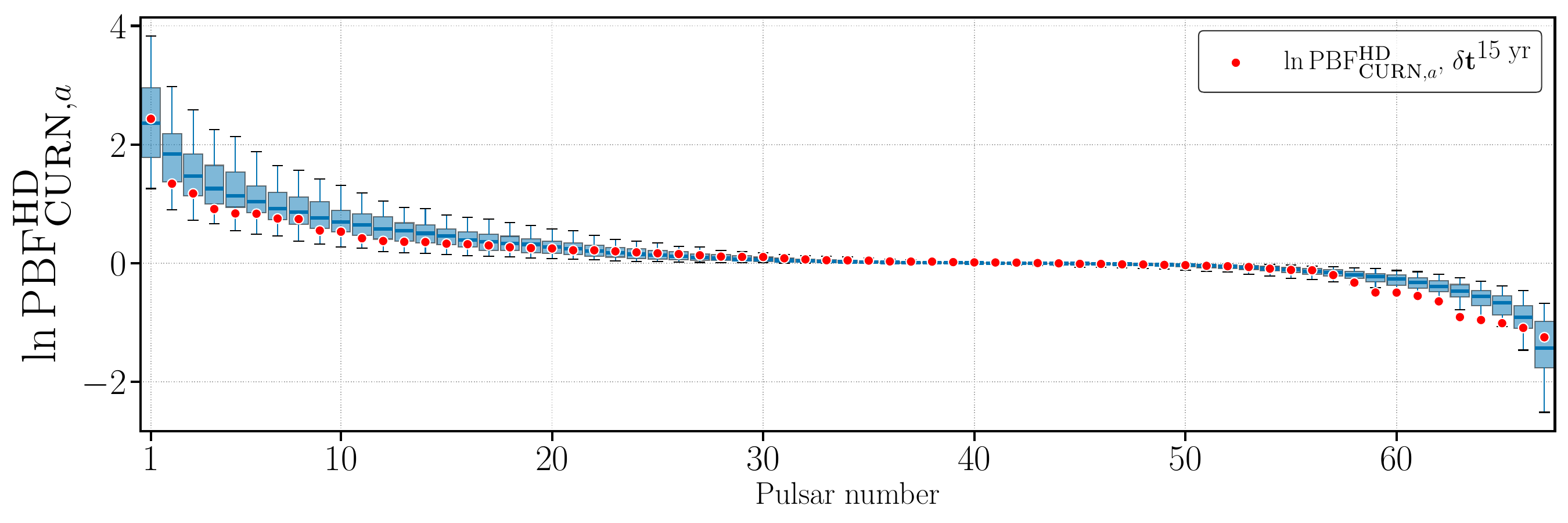}
    \caption{We compare results on 200 \hdposterior simulations to the results on $\tresidng$. The blue box and whisker plots correspond to the distribution of the pulsar with the $i^{\textrm{th}}$ highest value of $\ln\,\textrm{PBF}_{\textrm{\curn},a}^{\textrm{\hd}}$ for each simulation. For example, to construct the far left box we find the maximum $\ln\,\textrm{PBF}_{\textrm{\curn},a}^{\textrm{\hd}}$ for each simulation, and then build a distribution across simulations. For the far right box, we find the minimum $\ln\,\textrm{PBF}_{\textrm{\curn},a}^{\textrm{\hd}}$ for each simulation, and build a distribution, and so on. So the $\ln\,\textrm{PBF}_{\textrm{\curn},a}^{\textrm{\hd}}$ going into each blue box could be for a different pulsar for each simulation.}
    \label{fig:pbf_simulations_largest_to_largest}
\end{figure*}

We show the results of those simulations in Figs.~\ref{fig:pbf_simulation_results} and \ref{fig:pbf_simulations_largest_to_largest}.
In Fig.~\ref{fig:pbf_simulation_results}, we plot $\ln \textrm{PBF}_{\textrm{\curn},a}^{\textrm{\hd}}$ for each pulsar in red in the same order as Fig.~\ref{fig:dropout_results}. 
We show the distribution of $\ln \textrm{PBF}_{\textrm{\curn},a}^{\textrm{\hd}}$ for each pulsar across 200 \hdposterior simulations in the blue box and whisker plots. 
The boxes and whiskers correspond to the 50\% credible interval and the 5th and 95th percentiles respectively. 
The red points broadly agree with these distributions. 
The median simulated distribution for each pulsar falls above zero, corresponding to the \hd model being preferred. 
Calculating the percentile of the red point ($\ln \textrm{PBF}_{\textrm{\curn},a}^{\textrm{\hd}}$ on $\tresidng$) in each distribution yields a set of percentiles that are consistent with a uniform distribution between 0 and 1. 
This is what we expect if each pulsar is well-predicted by all of the others. 
In general, the pulsars with broader distributions and larger (positive or negative) values of $\ln \textrm{PBF}_{\textrm{\curn},a}^{\textrm{\hd}}$ correspond to the longest-timed and lowest-noise pulsars that have the greatest effect on the analysis.

In Fig.~\ref{fig:pbf_simulations_largest_to_largest} we present results from the same simulations, but we look at the distribution of the order statistics of $\ln \textrm{PBF}_{\textrm{\curn},a}^{\textrm{\hd}}$.
That is, the blue box and whisker to the furthest left corresponds to the distribution of the maximum $\ln \textrm{PBF}_{\textrm{\curn},a}^{\textrm{\hd}}$ across all pulsars for each simulation, so for each simulation we find the maximum $\ln \textrm{PBF}_{\textrm{\curn},a}^{\textrm{\hd}}$, and across simulations we build a distribution for that maximum. The second from left corresponds to the second largest $\ln \textrm{PBF}_{\textrm{\curn},a}^{\textrm{\hd}}$ in each simulation, the furthest to the right corresponds to the minimum value, and so on.  We see that our results are consistent with the simulations from the \hd model, and that in general we expect more pulsars to be better predicted by the \hd model. Crucially, simulations always result in a few pulsars that are better predicted by the \curn model, i.e., negative $\ln \textrm{PBF}_{\textrm{\curn},a}^{\textrm{\hd}}$. Therefore, negative $\ln\textrm{PBF}$ values are not immediately cause for concern as long as they are consistent with what we expect from simulations, which is the case here.

\section{\label{sec:waveforms}GW background waveforms}

The \hd model is preferred to the \curn model using the optimal statistic, Bayes factors, and PBFs. 
In this section, we show reconstructions of the \hd model compared to $\tresidng$. 
We also highlight covariances between different parts of the model to better understand the relationship between the GW background, the intrinsic pulsar noise, and the timing model for each pulsar. 
The figures presented in this section are meant to be representative and interpreted qualitatively to illustrate the contribution of different models and the covariances between those models; similar to the waveform reconstructions shown in Refs.~\cite{LIGOScientific:2016aoc,Lentati:2016ygu,Goncharov:2020krd}, for example. Similar figures have been shown before for noise models, e.g.~\cite{Lentati:2016ygu,Goncharov:2020krd,Larsen:2024vrt}, but not for the GW background model.

We first draw $\bvec^s \sim p(\bvec | \bm\Lambda^s, \tresid)$ using Eqs.~\eqref{eq:bvec_max_likelihood} and~\eqref{eq:bvec_covariance_matrix}, and then construct a model fit to the data by $\tresid^s = \tmat\bvec^s$ for each pulsar. 
As in the previous section, we separate the Gaussian process coefficients for intrinsic pulsar noise $\avec^s$, GW background $\avec_{\textrm{gw}}^s$, and timing model corrections $\epsvec^s$, which we use to inspect contributions from each part of the model independently. 

\begin{figure}
    \centering
    \includegraphics[width=\columnwidth]{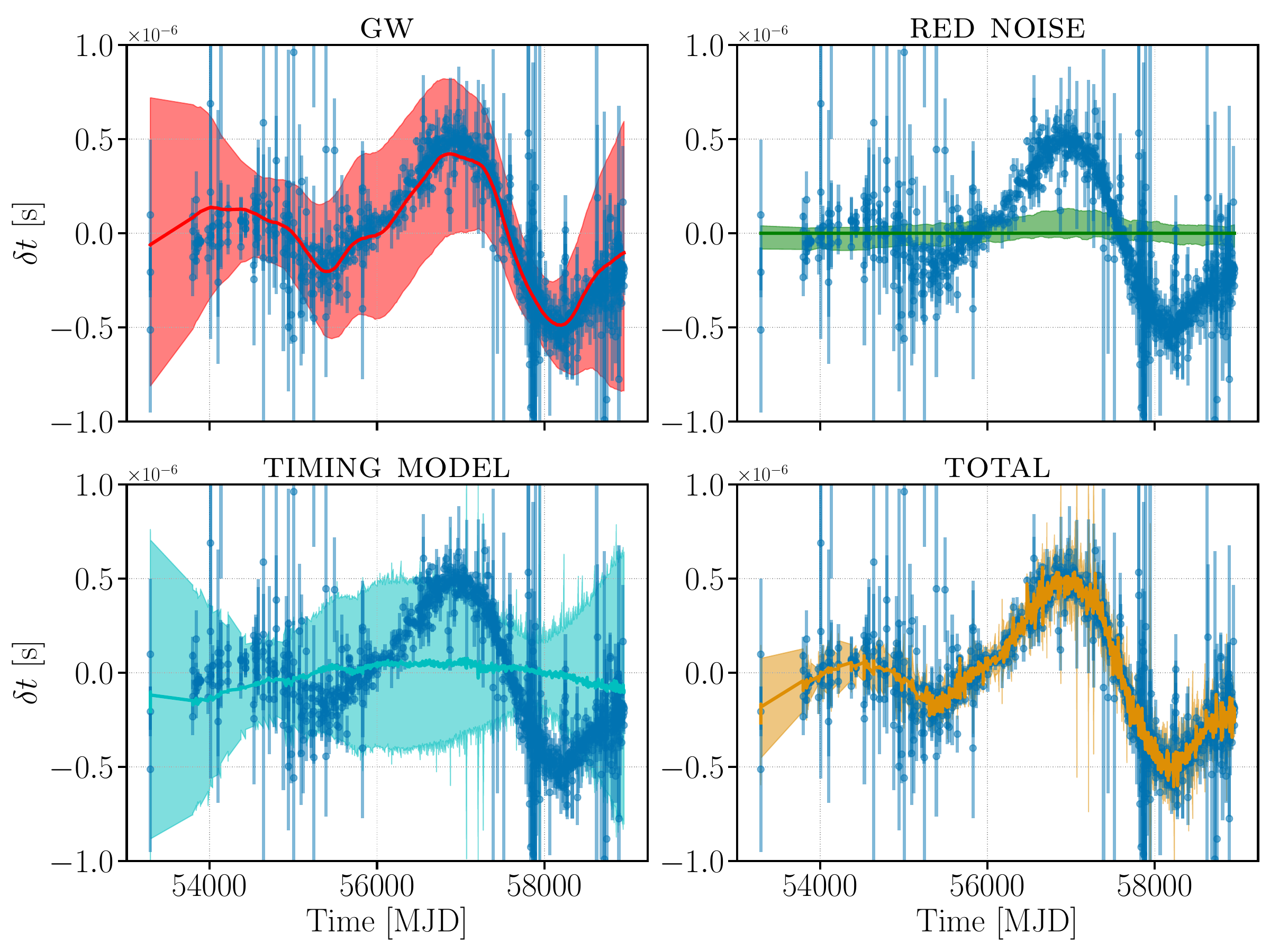}
    \caption{Waveform reconstruction for J1909-3744 (shaded regions) and timing residuals (blue dots). The solid line corresponds to the median reconstruction, and the shaded regions correspond to the 90\% credible interval. The blue points correspond to epoch-averaged residuals. In the top left is the contribution from the GW background, top right shows intrinsic pulsar noise, bottom left shows the timing model, and bottom right shows the combined total model. There is little evidence for intrinsic pulsar noise for this pulsar, and we can see that the frequency and spin-down components of the timing model (which give linear and quadratic offsets) are covariant with the lowest (and strongest) frequencies in the GW background. Regardless, the total model (bottom right) closely follows the data.}
    \label{fig:j1909_waveform}
\end{figure}

We show waveform reconstructions for pulsar J1909$-$3744 in Fig.~\ref{fig:j1909_waveform}. 
In each panel we show $\tresidng$ (averaged over day-long time-scales to reduce the number of points) and the contribution of one piece of our model.
For this pulsar, the total red noise is primarily due to GWs, indicated by the lack of intrinsic pulsar noise in the top right panel, and the fact that the GW background in the top left panel broadly follows $\tresidng$ plotted in blue. 
In the bottom left, we show the timing model in cyan. 
The spindown and spin frequency of the pulsar are covariant with the lowest frequencies of the GW background. 
This results in the broad uncertainties on the individual contributions from these models, but the narrow uncertainty on the combined contributions of all of the models in the bottom right (orange), which tracks the timing residuals closely. 

\begin{figure}
    \centering
    \includegraphics[width=\columnwidth]{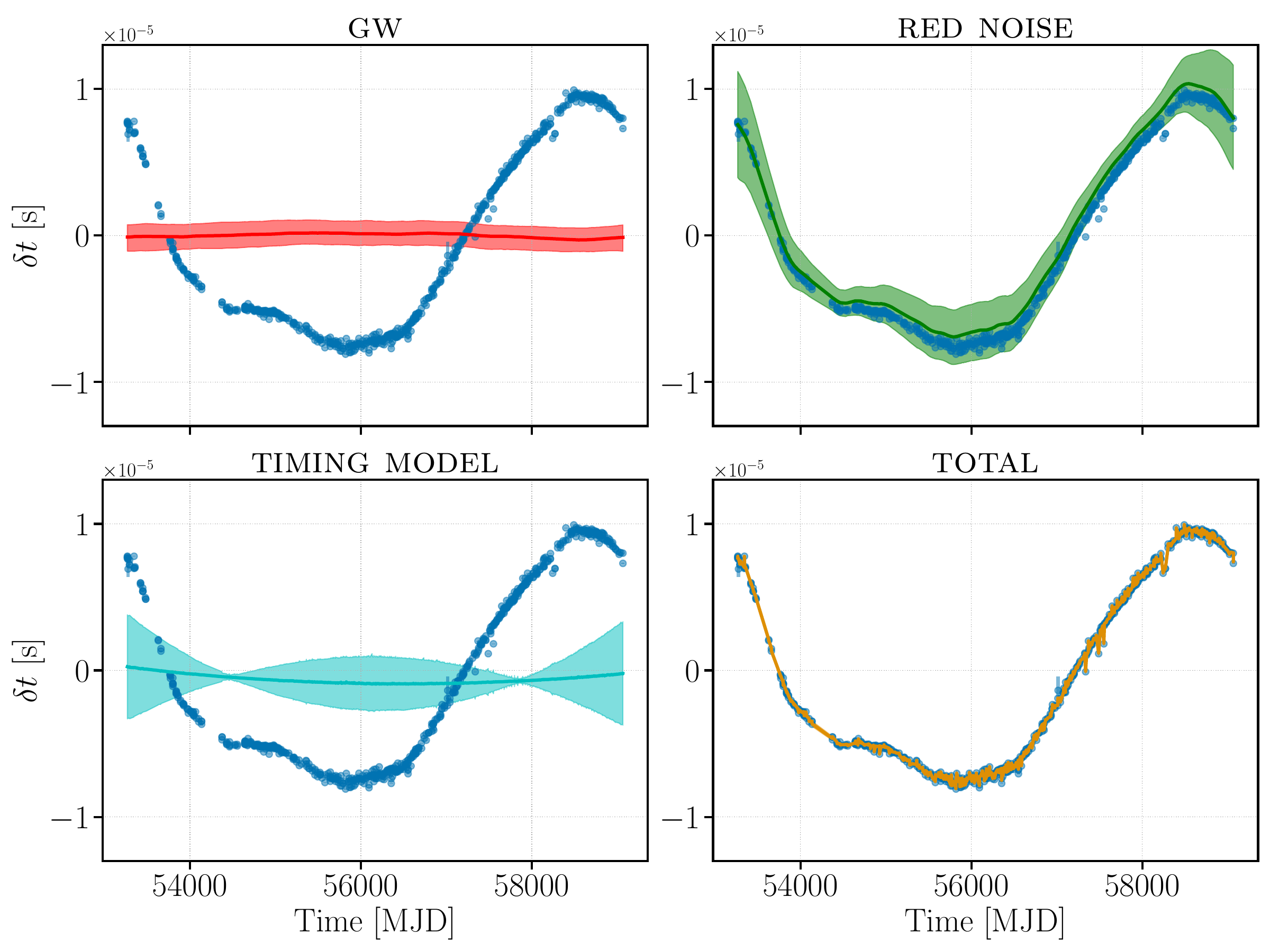}
    \caption{Waveform reconstruction for B1937+21 (shaded regions) and timing residuals (blue dots). In the top left is the contribution from the GW background, top right shows intrinsic pulsar noise, bottom left shows the timing model, and bottom right shows the combined total model. There is strong intrinsic pulsar noise in this pulsar, and in this case frequency and spin-down components of the timing model (which give linear and quadratic offsets) are covariant with the lowest (and strongest) frequencies in the intrinsic pulsar noise, while the GW background is significantly smaller than the noise. Again the total model closely tracks the data (bottom right).}
    \label{fig:b1937+21_waveform}
\end{figure}

We show a similar waveform reconstruction for pulsar B1937+21 in Fig.~\ref{fig:b1937+21_waveform}. 
The intrinsic pulsar noise dominates the pulsar's total red noise, as expected based on Fig.~\ref{fig:rn_comparison}.  
Waveform reconstructions for each pulsar are included as an online supplement. 
We find that in all cases, models represent reasonable fits to the data, which is expected based on the residual plots made with similar (single-pulsar) models in Ref.~\cite{NANOGrav:2023hde}. 
These figures are meant to illustrate the different contributions of each part of the model to the overall fit we make to each pulsar.

\section{\label{sec:conclusions}Conclusions}

The standard probabilistic model used to establish evidence for a GW background in Ref.~\cite{NANOGrav:2023gor} makes two assumptions motivated by theoretical expectations but also by computational convenience: that the background follows a power-law spectrum and that its inter-pulsar correlations conform to the Hellings--Downs pattern.
Deviations from these assumptions are expected from SMBHB astronomy and astrophysics, and in certain fundamental-physics scenarios, although it is unclear whether the deviations would be measurable in current data sets. 

In this paper, we examine the NANOGrav 15 yr data set~\cite{NANOGrav:2023hde} within the framework of Bayesian predictive model checking~\cite{ppc1,ppc2}, with the goal of testing the assumptions without comparison to alternative, more complex models.
The \textit{modus operandi} of Refs.~\cite{ppc1,ppc2} is that of using the fiducial model to simulate a population of replicated data sets from the real-data parameter posteriors, and then comparing the values of multiple statistics of interest in real data and across the replications.

The optimal statistic~\cite{chamberlinOS,Vigeland:2018ipb} was used in Ref.~\cite{NANOGrav:2023gor} to establish the presence of inter-pulsar timing-residual correlations.  
Within the replication framework, we can account fully for the dependence of the optimal statistic on the uncertain noise parameters~\cite{ppc1}, building a Bayesian $p$-value that falsifies the no-correlation hypothesis at the 3.2$\,\sigma$ level for the NANOGrav data set. 
That is, we find that data replications obtained from a spatially uncorrelated model can rarely reproduce the value of the optimal statistic seen for real data.
The Bayesian $p$-value is averaged over the noise-parameter posterior, accounting fairly for the overall risk of false rejection.
If instead we build our replications from the Hellings--Downs model, we find a $p$-value $\sim 0.5$, as expected if that model is correct.
We also find no anomalies when we use optimal-statistic variants built to be sensitive to monopolar or dipolar correlations.

Moving on from the frequentist flavor of this optimal-statistic analysis to Bayesian model comparison, we evaluate the relative predictive performance of the Hellings--Downs and spatially uncorrelated models by way of the leave-one-out cross-validation pseudo Bayes factor~\cite{ppc2}. We find that the Hellings--Downs model is favored at the 3$\,\sigma$ level. That is, we find that data replications obtained from a spatially uncorrelated model can rarely reproduce the pseudo Bayes factor seen for real data.
We also verify that the binned correlation coefficients estimated from real data are consistent with the distribution expected under the Hellings--Downs hypothesis. 
Altogether, we find that the 15 yr NANOGrav data set is consistent with the hypothesis of Hellings--Downs correlations, with no evidence for alternative correlation patterns.

We test the assumption that the GW background has a power-law spectrum by comparing the real-data posteriors of the spectral coefficients (i.e., the root-mean-square Fourier amplitudes at each frequency) with their distribution across replicated data sets.
Although some ``spikes'' are evident in the spectral plots, we find that they are not statistically significant---they are not unlikely in the replicated population.
As a byproduct of this analysis, Fourier-amplitude posteriors provide a probabilistic reconstruction of the putative GW signal, as seen most strikingly in Fig.~\ref{fig:j1909_waveform} for pulsar J1909$-$3744.

This paper details an extensive but certainly not exhaustive reanalysis of the NANOGrav 15 yr data set.
Our overall finding is that the data are consistent with a simple power-law GW background with isotropic Hellings--Downs correlations.
Future more expansive and sensitive data sets will require more sophisticated data models; the framework introduced in Refs.~\cite{ppc1,ppc2} and exemplified here can tell us when we have reached that threshold.

\section*{Authorship contributions}
This paper uses a decade's worth of pulsar timing observations and is the product of the work of many people. P.M.M. helped conceive the project, wrote and developed code to perform the analysis, created all figures, and wrote and edited the text. M.V. helped conceive the project, developed code, performed parts of the analysis, ran preliminary analyses, and helped write and edit the text. K.Ch. helped conceive the project, guided direction of the analysis, and wrote and edited the text. B.L.,  S.V., T.D., and D.R.S., gave constructive comments that improved the manuscript, as did members of the NANOGrav Detection Working Group.

G.A., A.A, A.M.A., Z.A., P.T.B., P.R.B., H.T.C.,
K.C., M.E.D, P.B.D., T.D., E.C.F, W.F., E.F., G.E.F.,
N.G.D., D.C.G., P.A.G., J.G., R.J.J., M.L.J., D.L.K.,
M.K., M.T.L., D.R.L., J.L., R.S.L., A.M., M.A.M.,
N.M., B.W.M., C.N., D.J.N., T.T.N., B.B.P.P., N.S.P.,
H.A.R., S.M.R., P.S.R., A.S., C.S., B.J.S.A., I.H.S.,
K.S., A.S., J.K.S., and H.M.W. developed timing models and ran observations for the NANOGrav 15 yr data
set.

\section*{Acknowledgements}

The authors thank Rutger van Haasteren and two anonymous referees for their constructive comments on the manuscript.

The NANOGrav Collaboration receives support from National Science Foundation (NSF) Physics Frontiers Center award Nos. 1430284 and 2020265, the Gordon and Betty Moore Foundation, NSF AccelNet award No. 2114721, an NSERC Discovery Grant, and CIFAR. The Arecibo Observatory is a facility of the NSF operated under cooperative agreement (AST-1744119) by the University of Central Florida (UCF) in alliance with Universidad Ana G. Méndez (UAGM) and Yang Enterprises (YEI), Inc. The Green Bank Observatory is a facility of the NSF operated under cooperative agreement by Associated Universities, Inc. The National Radio Astronomy Observatory is a facility of the NSF operated under cooperative agreement by Associated Universities, Inc.
Part of this research was performed at the Jet Propulsion Laboratory, under contract with the National Aeronautics and Space Administration.
Copyright 2024.

L.B. acknowledges support from the National Science Foundation under award AST-1909933 and from the Research Corporation for Science Advancement under Cottrell Scholar Award No. 27553.
P.R.B. is supported by the Science and Technology Facilities Council, grant number ST/W000946/1.
S.B. gratefully acknowledges the support of a Sloan Fellowship, and the support of NSF under award \#1815664.
M.C. and S.R.T. acknowledge support from NSF AST-2007993.
M.C. and N.S.P. were supported by the Vanderbilt Initiative in Data Intensive Astrophysics (VIDA) Fellowship.
K.Ch., A.D.J., and M.V. acknowledge support from the Caltech and Jet Propulsion Laboratory President's and Director's Research and Development Fund.
K.Ch. and A.D.J. acknowledge support from the Sloan Foundation.
Support for this work was provided by the NSF through the Grote Reber Fellowship Program administered by Associated Universities, Inc./National Radio Astronomy Observatory.
Pulsar research at UBC is supported by an NSERC Discovery Grant and by CIFAR.
K.Cr. is supported by a UBC Four Year Fellowship (6456).
M.E.D. acknowledges support from the Naval Research Laboratory by NASA under contract S-15633Y.
T.D. and M.T.L. are supported by an NSF Astronomy and Astrophysics Grant (AAG) award number 2009468.
E.C.F. is supported by NASA under award number 80GSFC21M0002.
G.E.F., S.C.S., and S.J.V. are supported by NSF award PHY-2011772.
K.A.G. and S.R.T. acknowledge support from an NSF CAREER award \#2146016.
The work of N.La., X.S., and D.W. is partly supported by the George and Hannah Bolinger Memorial Fund in the College of Science at Oregon State University.
N.La. acknowledges the support from Larry W. Martin and Joyce B. O'Neill Endowed Fellowship in the College of Science at Oregon State University.
Part of this research was carried out at the Jet Propulsion Laboratory, California Institute of Technology, under a contract with the National Aeronautics and Space Administration (80NM0018D0004).
D.R.L. and M.A.M. are supported by NSF \#1458952.
M.A.M. is supported by NSF \#2009425.
C.M.F.M. was supported in part by the National Science Foundation under Grants No. NSF PHY-1748958 and AST-2106552.
A.Mi. is supported by the Deutsche Forschungsgemeinschaft under Germany's Excellence Strategy - EXC 2121 Quantum Universe - 390833306.
The Dunlap Institute is funded by an endowment established by the David Dunlap family and the University of Toronto.
K.D.O. was supported in part by NSF Grant No. 2207267.
T.T.P. acknowledges support from the Extragalactic Astrophysics Research Group at E\"{o}tv\"{o}s Lor\'{a}nd University, funded by the E\"{o}tv\"{o}s Lor\'{a}nd Research Network (ELKH), which was used during the development of this research.
H.A.R. is supported by NSF Partnerships for Research and Education in Physics (PREP) award No. 2216793.
S.M.R. and I.H.S. are CIFAR Fellows.
Portions of this work performed at NRL were supported by ONR 6.1 basic research funding.
J.D.R. also acknowledges support from start-up funds from Texas Tech University.
J.S. is supported by an NSF Astronomy and Astrophysics Postdoctoral Fellowship under award AST-2202388, and acknowledges previous support by the NSF under award 1847938.
C.U. acknowledges support from BGU (Kreitman fellowship), and the Council for Higher Education and Israel Academy of Sciences and Humanities (Excellence fellowship).
C.A.W. acknowledges support from CIERA, the Adler Planetarium, and the Brinson Foundation through a CIERA-Adler postdoctoral fellowship.
O.Y. is supported by the National Science Foundation Graduate Research Fellowship under Grant No. DGE-2139292.

\bibliography{main}
\end{document}